\documentclass[cernpreprint,texlive=2016,backref=false,UKenglish,texmf]{atlasdoc}
 
\usepackage[subcaption]{atlaspackage}
\usepackage{atlasbiblatex}
\usepackage[BSM]{atlasphysics}
 
\usepackage{multirow}
\newcolumntype{H}{>{\setbox0=\hbox\bgroup}c<{\egroup}@{}}
 
\graphicspath{{logos/}{figures/}}
 
\usepackage{ANA-SUSY-2018-21-PAPER-defs}

\AtlasTitle{Search for top squarks in events with a Higgs or $Z$ boson using 139~fb$^{-1}$ of $pp$ collision data at $\sqrt{s}=13$~TeV with the ATLAS detector}
 
\PreprintIdNumber{CERN-EP-2020-074}
 
\AtlasAbstract{
This paper presents a search for direct top squark pair production in events with missing transverse momentum plus either a pair of jets consistent with Standard Model Higgs boson decay into $b$-quarks or a same-flavour opposite-sign dilepton pair with an invariant mass consistent with a $Z$ boson. The analysis is performed using the proton--proton collision data at $\sqrt{s}=13$ TeV collected with the ATLAS detector during the LHC Run-2, corresponding to an integrated luminosity of 139~fb$^{-1}$. No excess is observed in the data above the Standard Model predictions.
The results are interpreted in simplified models featuring direct production of pairs of either the lighter top squark ($\tilde{t}_1$) or the heavier top squark ($\tilde{t}_2$), excluding at 95\% confidence level $\tilde{t}_1$ and $\tilde{t}_2$ masses up to about 1220 and 875~GeV, respectively.
}
 
\author{The ATLAS Collaboration}
 
\AtlasDate{\today}
 
\AtlasJournal{Eur. Phys. J. C}
 
\AtlasJournalRef{Eur. Phys. J. C 80 (2020) 1080}
 
\AtlasDOI{10.1140/epjc/s10052-020-08469-8}
\hypersetup{pdftitle={ATLAS document},pdfauthor={The ATLAS Collaboration}}
 
\begin{document}
 
\maketitle
 
\section{Introduction}
\label{sec:intro}
 
Supersymmetry (SUSY)~\cite{Golfand:1971iw,Volkov:1973ix,Wess:1974tw,Wess:1974jb,Ferrara:1974pu,Salam:1974ig} is one of the most studied frameworks to extend the Standard Model (SM) beyond the electroweak scale.
It predicts new bosonic (fermionic) partners for the known fermions (bosons).
Assuming $R$-parity conservation~\cite{Farrar:1978xj}, SUSY particles are produced in pairs and the lightest supersymmetric particle (LSP) is stable, providing a possible dark-matter candidate. The SUSY partners of the Higgs bosons and electroweak gauge bosons mix to form the mass eigenstates known as charginos ($\tilde{\chi}^{\pm}_{k}$, $k = 1, 2$) and neutralinos ($\tilde{\chi}^{0}_{m}$, $m = 1, 2, 3, 4$), where the increasing index denotes increasing mass.
The scalar partners of right-handed and left-handed quarks, \squarkR and \squarkL squarks, mix to form two mass eigenstates, $\tilde{q}_{1}$ and $\tilde{q}_{2}$, with $\tilde{q}_{1}$ defined to be the lighter of the two.
To address the SM hierarchy problem~\cite{Sakai:1981gr,Dimopoulos:1981yj,Ibanez:1981yh,Dimopoulos:1981zb}, \TeV-scale masses are favoured~\cite{Barbieri:1987fn,deCarlos:1993yy} for the supersymmetric partners of the gluons, and the top squarks~\cite{Inoue:1982pi,Ellis:1983ed}.
 
Top squark production with SM Higgs ($h$) or $Z$ bosons in the decay chain can appear either in production of the lighter top squark mass eigenstate (\stopone) decaying via $\stopone\to t\ninotwo$ with $\ninotwo\to h/Z\ninoone$,
or in production of the heavier top squark mass eigenstate (\stoptwo) decaying via $\stoptwo\to Z \stopone$ with $\stopone \to t^{(*)} \ninoone$, as illustrated in Figure~\ref{fig:susyfeynman}.
Unlike other top squark models, these signals can be efficiently distinguished from the SM top quark pair production ($\ttbar$) background by requiring either a same-flavour opposite-sign (SF-OS) lepton pair originating from the $Z\to\ell^+\ell^-$ ($\ell \equiv e, \mu$) decay or a pair of $b$-tagged jets originating from the $h\to\bbbar$ decay,
plus the presence of an additional lepton produced in the decay of the top quarks in the event.
 
Simplified models~\cite{Alwall:2008ve,Alwall:2008ag,Alves:2011wf} are used for the analysis optimisation and interpretation of the results.
In these models, direct top squark pair production is considered and all SUSY particles are decoupled except for the top squarks and the neutralinos involved in their decay. In all cases the $\ninoone$ is assumed to be the LSP.
Simplified models featuring direct $\stopone$ production with $\stopone \to t\ninotwo$ and decays via either Higgs ($\ninotwo \to h\ninoone$)  or $Z$ ($\ninotwo \to Z\ninoone$) bosons with different branching ratio values are considered. In these models, the \ninoone is assumed to be very light and the $\ninotwo - \ninoone$ mass difference to be large enough to allow on-shell Higgs or $Z$ boson decays.
 
Additional simplified models featuring direct $\stoptwo$ production with $\stoptwo\to Z\stopone$ decays and $\stopone \to t^{(*)}\ninoone$ are also considered.
The mass difference between the $\stopone$ and $\ninoone$ is set to be smaller than the $W$ boson mass, and the four-body decay $\stopone \to b ff'\ninoone$ is assumed to occur, where $f$ and $f'$ are two fermions from the $W^*$ decay, such as $\stopone \to b \ell\nu\ninoone$.
Direct production of \stopone\ pairs is not considered in the \stoptwo\ production simplified models. The \stopone\ pair production contribution to the selections presented in this paper has been found to be negligible.
 
This paper presents the results of a search for top squarks in final states with Higgs or $Z$ bosons at $\sqrt{s}=13$~\TeV\ using the complete data sample collected with the ATLAS detector~\cite{PERF-2007-01} in proton--proton ($pp$) collisions during Run-2 of the LHC (2015--2018), corresponding to 139~\ifb.
Searches for top squark production in events involving Higgs or $Z$ bosons have been performed previously by both ATLAS~\cite{SUSY-2016-20,Stop2_Run1} and CMS~\cite{CMS-SUS-13-024,Stop2_Run1cms}.
 
\begin{figure}[htb]
\begin{center}
\begin{subfigure}[t]{0.48\textwidth}
\centering
\includegraphics[width=0.8\textwidth]{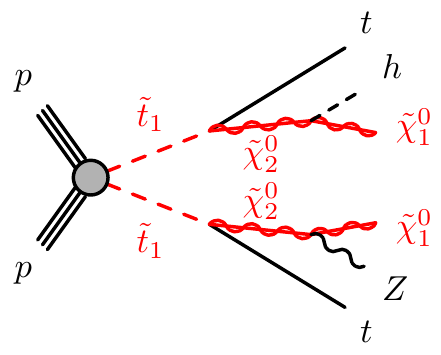}
\caption{}
\end{subfigure}
\begin{subfigure}[t]{0.48\textwidth}
\centering
\includegraphics[width=0.8\textwidth]{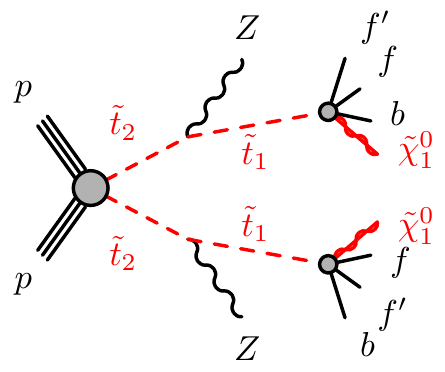}
\caption{}
\end{subfigure}
\end{center}
\caption{Diagrams for the top squark pair production processes considered in this analysis: (a) $\stopone \to t \ninotwo$ with $\ninotwo\to h/Z\ninoone$ decays (showing for illustration the case where the two $\ninotwo$ decay differently, although events with the same $\ninotwo$ decays are also considered in the analysis), and (b) $\stoptwo \to Z \stopone$ with $\stopone \to b ff'\ninoone$ decays.}
\label{fig:susyfeynman}
\end{figure}

\section{ATLAS detector}
\label{sec:detector}
 
\newcommand{\AtlasCoordFootnote}{
ATLAS uses a right-handed coordinate system with its origin at the nominal interaction point (IP)
in the centre of the detector and the \(z\)-axis along the beam pipe.
The \(x\)-axis points from the IP to the centre of the LHC ring,
and the \(y\)-axis points upwards.
Cylindrical coordinates \((r,\phi)\) are used in the transverse plane,
\(\phi\) being the azimuthal angle around the \(z\)-axis.
The pseudorapidity is defined in terms of the polar angle \(\theta\) as \(\eta = -\ln \tan(\theta/2)\),
and the rapidity is defined as \(y = (1/2) [(E+p_z)/(E-p_z)]\).
}
 
The ATLAS detector~\cite{PERF-2007-01,PIX-2018-001} at the LHC is a multipurpose particle detector with a forward--backward symmetric cylindrical geometry and a near \(4\pi\) coverage in solid angle.\footnote{\AtlasCoordFootnote}
It consists of an inner tracking detector (ID) surrounded by a thin superconducting solenoid providing a \SI{2}{\tesla} axial magnetic field, electromagnetic (EM) and hadron calorimeters, and a muon spectrometer (MS).
The inner tracking detector covers the pseudorapidity range \(|\eta| < 2.5\).
It consists of silicon pixel, silicon microstrip, and transition radiation tracking detectors.
Lead/liquid-argon (LAr) sampling calorimeters provide EM energy measurements with high granularity.
A steel/scintillator-tile hadron calorimeter covers the central pseudorapidity range \(|\eta| < 1.7\).
The endcap and forward regions are instrumented with LAr calorimeters for both the EM and hadronic energy measurements up to \(|\eta| = 4.9\).
The muon spectrometer surrounds the calorimeters and is based on three large air-core toroidal superconducting magnets with eight coils each.
The field integral of the toroids ranges between \num{2} and \SI{6}{\tesla\metre} across most of the detector.
The muon spectrometer includes a system of precision tracking chambers and fast detectors for triggering.
A two-level trigger system is used to select events~\cite{TRIG-2016-01}.
The first-level trigger is implemented in hardware and uses a subset of the detector information
to reduce the accepted rate to at most \SI{100}{\kHz}.
This is followed by a software-based trigger that
reduces the accepted event rate to \SI{1}{\kHz} on average
depending on the data-taking conditions.
 
\section{Data set and simulated event samples}
\label{sec:dataMC}
 
The data were collected by the ATLAS detector during the LHC Run-2 (2015--2018) with a peak instantaneous luminosity of ${\cal L}=2.1\times 10^{34}$~cm$^{-2}$s$^{-1}$, resulting in a mean number of $pp$ interactions per bunch crossing of $\langle \mu \rangle = 34$.
Data quality requirements are applied to ensure that all subdetectors were operating normally, and that the LHC beams were in stable-collision mode.
The integrated luminosity of the resulting data sample is 139~\ifb.
The uncertainty in the combined 2015--2018 integrated luminosity is 1.7\%. It is derived from the calibration of the luminosity scale using $x$--$y$ beam-separation scans, following a methodology similar to that detailed in Ref.~\cite{DAPR-2013-01}, and using the LUCID-2 detector for the baseline luminosity measurements~\cite{LUCID2}.
 
Monte Carlo (MC) simulated event samples are used to aid the estimation of the background from SM processes and to model the SUSY signal.
The choices of MC event generator, parton shower and hadronisation, the cross-section normalisation, the parton distribution function (PDF) set and the set of tuned parameters (tune) for the underlying event of these samples are summarised in Table~\ref{tab:MC}. More details of the event generator configurations can be found in Refs.~\cite{ATL-PHYS-PUB-2016-004,ATL-PHYS-PUB-2016-003,ATL-PHYS-PUB-2016-002,ATL-PHYS-PUB-2016-005}.
Cross-sections calculated at next-to-next-to-leading order (NNLO) in quantum chromodynamics (QCD) including resummation of next-to-next-to-leading logarithmic (NNLL) soft-gluon terms were used for top quark production processes.
For production of top quark pairs in association with vector or Higgs bosons, cross-sections calculated at next-to-leading order (NLO) were used, and the event generator NLO cross-sections from \SHERPA~\cite{gleisberg:2008ta} were used when normalising the multi-boson backgrounds.
In all MC samples, except those produced by \SHERPA, the {\textsc EvtGen}~v1.2.0 program~\cite{EvtGen} was used to model the properties of the bottom and charm hadron decays.

\begin{table*}[t]
\caption{Simulated signal and background event samples: the corresponding event generator used for the hard-scatter process, the generator used to model the parton showering, the source of the cross-section used for normalisation, the PDF set and the underlying-event tune are shown.\vspace{-0.8cm}}
\label{tab:MC}
\begin{center}
\resizebox{\textwidth}{!}
{\small
\begin{tabular}{l l c c c c}
\hline
Physics process    & Generator & Parton shower & Cross-section & PDF set & Tune \\
&	      & 	      & normalisation & 	&      \\
\hline
SUSY Signals    	   & \AMCATNLO 2.6.2~\cite{Alwall:2014hca} & \PYTHIA 8.212~\cite{Sjostrand:2014zea} & NNLO+NNLL~\cite{Beenakker:2016lwe,Beenakker:1997ut,Beenakker:2010nq,Beenakker:2016gmf} & NNPDF2.3LO~\cite{Ball:2012cx} & A14~\cite{ATL-PHYS-PUB-2014-021} \\
\hline
$t\bar{t} Z/\gamma^{*}$, $\ttbar W$ & \AMCATNLO 2.3.3   & \PYTHIA 8.210	& NLO~\cite{Alwall:2014hca}   & NNPDF2.3LO & A14    \\
Diboson    	   & \SHERPA 2.2.2~\cite{gleisberg:2008ta}     & \SHERPA 2.2.2	& Generator NLO	 & NNPDF3.0NNLO~\cite{Ball:2014uwa}	     & \SHERPA default \\
$\ttbar$ & \POWHEGBOXV{v2}~\cite{Frixione:2007vw}  &  \PYTHIA 8.230 & NNLO+NNLL~\cite{Czakon:2013goa,Czakon:2012pz,Czakon:2012zr,Baernreuther:2012ws,Cacciari:2011hy,Czakon:2011xx}  & NNPDF3.0NLO & A14  \\
\hline
$t\bar{t}h$	   & \POWHEGBOXV{v2}      & \PYTHIA 8.230          & NLO~\cite{Dittmaier:2012vm}   & NNPDF2.3LO  & A14  \\
Single-top, $Wt$ & \POWHEGBOXV{v2}  & \PYTHIA 8.230 &NNLO+NNLL~\cite{Kidonakis:2011wy,Kidonakis:2010ux,Kidonakis:2010tc} & NNPDF3.0NLO & A14  \\
 
$Wh$, $Zh$     & \PYTHIA 8.186~\cite{Sjostrand:2007gs}       & \PYTHIA 8.186 & NLO~\cite{Dittmaier:2012vm}   & NNPDF2.3LO     & A14  \\
$t\bar{t} WW$, $\ttbar\ttbar$  & \AMCATNLO 2.2.2       & \PYTHIA 8.186   & NLO~\cite{Alwall:2014hca}	 & NNPDF2.3LO	  & A14  \\
$t\bar{t}t$            & \AMCATNLO 2.2.2       & \PYTHIA 8.186   & Generator LO	 & NNPDF2.3LO	  & A14  \\
$tZ$                   & \AMCATNLO 2.3.3       & \PYTHIA 8.186   & Generator LO                   	    & NNPDF2.3LO     & A14  \\
$tWZ$                  & \AMCATNLO 2.3.3       & \PYTHIA 8.212   & Generator NLO  & NNPDF2.3LO     & A14  \\
Triboson	   & \SHERPA 2.2.2        & \SHERPA 2.2.2    & Generator NLO            & NNPDF3.0NNLO    & \SHERPA default \\
\hline
\end{tabular}
}
\end{center}
\end{table*}

SUSY signal samples were generated with {\scshape MG5\_aMC\/@NLO} 2.6.2~\cite{Alwall:2014hca} interfaced to \PYTHIA 8.212~\cite{Sjostrand:2014zea} for the parton showering (PS) and hadronisation.
The matrix element (ME) calculation was performed at tree level and includes the emission of up to two additional partons for all signal samples.
\MADSPIN~\cite{Artoisenet:2012st} was used to model the $\stopone \rightarrow b f f' \ninoone$ decays. \MADSPIN emulates kinematic distributions to a good approximation without calculating the full ME.
The PDF set used for the generation of the signal samples was NNPDF2.3LO~\cite{Ball:2012cx} with the A14~\cite{ATL-PHYS-PUB-2014-021} set of tuned underlying-event and shower parameters (UE tune). The ME--PS matching was performed with the CKKW-L prescription~\cite{Lonnblad:2011xx}, with a matching scale set to one quarter of the top squark mass.
All signal cross-sections were calculated to approximate NNLO in the strong coupling constant, adding the resummation of soft gluon emission at NNLL accuracy (approximate NNLO+NNLL)~\cite{Beenakker:2016lwe,Beenakker:1997ut,Beenakker:2010nq,Beenakker:2016gmf}. The nominal cross-section and its uncertainty were derived using the PDF4LHC15\_mc PDF set, following the recommendations of Ref.~\cite{Butterworth:2015oua}.
 
To simulate the effects of additional $pp$ collisions in the same and nearby bunch crossings (pile-up), additional interactions were generated using the soft QCD processes provided by \PYTHIA 8.186 with the A3 tune~\cite{ATL-PHYS-PUB-2016-017} and the MSTW2008LO PDF set~\cite{Martin:2009iq}, and overlaid onto each simulated hard-scatter event.
The MC samples were reweighted so that the pile-up distribution matches the one observed in the data. The MC samples were processed through an ATLAS detector simulation~\cite{Aad:2010ah} based on \GEANT 4~\cite{Agostinelli:2002hh} or, in the case of $\ttbar t$ and the SUSY signal samples, a fast simulation using a parameterisation of the calorimeter response and \GEANT 4 for the other parts of the detector. All MC samples were reconstructed in the same manner as the data.
 
\section{Event selection}
\label{sec:selection}
 
Candidate events are required to have a reconstructed vertex~\cite{ATL-PHYS-PUB-2015-026} with at least two associated tracks with transverse momentum (\pT) larger than 500~\MeV\ that are consistent with originating from the beam collision region in the $x$--$y$ plane.
The primary vertex in the event is the vertex with the highest sum of squared transverse momenta of associated tracks.
 
Two categories of leptons (electrons and muons) are defined: `candidate' and `signal' (the latter being a subset of the `candidate' leptons satisfying tighter selection criteria).
Electron candidates are reconstructed from isolated electromagnetic calorimeter energy deposits matched to ID tracks and are required to have $|\eta|<2.47$, a transverse momentum $\pT>\SI{4.5}{\GeV}$, and to satisfy the `LooseAndBLayer' requirement defined in Ref.~\cite{PERF-2017-01}, which is based on a likelihood using measurements of shower shapes in the calorimeter and track properties in the ID as input variables.
 
Muon candidates are reconstructed in the region $|\eta|<2.4$ from MS tracks matching ID tracks.
Candidate muons are required to have $\pT>\SI{4}{\GeV}$ and satisfy the `medium' identification requirements defined in Ref.~\cite{PERF-2015-10}, based on the number of hits in the different ID and MS subsystems, and 
on the ratio  of the charge and momentum ($q/p$) measured in the ID and MS divided by the sum in quadrature of their corresponding uncertainties.
 
The tracks associated with the lepton candidates are required to have a significance of the transverse impact parameter relative to the reconstructed primary vertex, $d_0$, of $\vert d_0\vert/\sigma(d_0) < 5$ for electrons and $\vert d_0\vert/\sigma(d_0) < 3$ for muons, and a longitudinal impact parameter relative to the reconstructed primary vertex, $z_0$, satisfying $\vert z_0 \sin\theta\vert < 0.5$~mm.
 
Jets are reconstructed from three-dimensional energy clusters in the calorimeter~\cite{ATL-PHYS-PUB-2015-036} using the anti-$k_t$ jet clustering algorithm~\cite{Cacciari:2008} with a radius parameter $R=0.4$. Only jet candidates with $\pT>\SI{20}{\GeV}$ and $|\eta|<2.8$ are considered.
Jets are calibrated using MC simulation with corrections obtained from in situ techniques~\cite{PERF-2016-04}.
To reduce the effects of pile-up, jets with $\pt<\SI{120}{\GeV}$ and $|\eta|<2.5$ are required to have a significant fraction of their associated tracks compatible with originating from the primary vertex, as defined by the jet vertex tagger~\cite{PERF-2014-03}.
This requirement reduces the fraction of jets from pile-up to 1\%, with an efficiency for hard-scatter jets of about 90\%.
Events are discarded if they contain any jet with $\pT>\SI{20}{\GeV}$ not satisfying basic quality selection criteria designed to reject detector noise and non-collision backgrounds~\cite{ATLAS-CONF-2015-029}.
 
Identification of jets containing $b$-hadrons ($b$-tagging) is performed with 
a multivariate discriminant that makes use of track impact parameters and reconstructed secondary vertices~\cite{ATL-PHYS-PUB-2017-013, FTAG-2018-01}.
Jets are considered as $b$-tagged if they fulfil a requirement corresponding to a 77\% average efficiency obtained for jets containing $b$-hadrons in simulated $\ttbar$ events.
The rejection factors for light-quark and gluon jets, jets containing $c$-hadrons and jets containing hadronically decaying $\tau$-leptons in simulated $\ttbar$ events are approximately 113, 16 and 4, respectively.
 
Jet candidates with $\pT<200$~\GeV\ within an angular distance $\Delta R =\sqrt{(\Delta y)^2+(\Delta\phi)^2} = 0.2$ of an electron candidate are discarded, unless the jet has a value of the $b$-tagging discriminant larger than the value corresponding to approximately 85\% $b$-tagging efficiency, in which case the lepton is discarded since it is likely to have originated from a semileptonic $b$-hadron decay. The same procedure is applied to jets within $\Delta R = 0.2$ of a muon candidate irrespective of the jet \pT.
Any remaining electron candidate within $\Delta R=0.4$ of a non-pile-up jet, and any muon candidate within $\Delta R= \operatorname{min}\{0.4, 0.04 + \pt(\mu)/\SI{10}{\GeV}\}$ of a non-pile-up jet is discarded.
In the latter case, if the jet has fewer than three associated tracks, the muon is retained and the jet is discarded instead
to avoid inefficiencies for high-energy muons undergoing significant energy loss in the calorimeter.
Finally, any electron candidate sharing an ID track with a remaining muon candidate is also removed.
 
Tighter requirements on the lepton candidates are imposed, which are then referred to as `signal' electrons or muons.
Signal electrons must satisfy the `Medium' identification requirement as defined in Ref.~\cite{PERF-2017-01} and signal muons must have $\pt>5$~\GeV.
Isolation requirements are applied to both the signal electrons and muons.
The  scalar sum of the \pt of tracks within a variable-size cone around the lepton, excluding its own track, must be less than 6\% of the lepton \pt; these tracks are required to be associated with the primary vertex to limit sensitivity to pile-up.
The size of the track isolation cone for electrons (muons) is given by the smaller of $\Delta R = \SI{10}{\GeV}/\pt$ and $\Delta R = 0.2\,(0.3)$.
In addition, in the case of electrons the energy of calorimeter energy clusters in a cone of $\Delta R_\eta = \sqrt{(\Delta\eta)^2+(\Delta\phi)^2} = 0.2$ around the electron
(excluding the deposition from the electron itself) must be less than 6\% of the electron \pt.
 
Simulated events are corrected for differences between data and MC simulation in jet vertex tagger and $b$-tagging efficiencies as well as $b$-tagging mis-tag rates~\cite{PERF-2014-03,FTAG-2018-01, ATLAS-CONF-2018-001, ATLAS-CONF-2018-006}. Corrections are also applied to account for minor differences between data and MC simulation in the signal-lepton trigger, reconstruction, identification and isolation efficiencies.
 
The missing transverse momentum vector, whose magnitude is denoted by \MET, is defined as the negative vector sum of the transverse momenta of all identified electrons, muons and jets, plus an additional soft term. The soft term is constructed from all tracks that originate from the primary vertex but are not associated with any identified lepton or jet. In this way, the \MET is adjusted for the best calibration of leptons and jets, while contributions from pile-up interactions are suppressed through the soft term~\cite{ATL-PHYS-PUB-2015-027,ATLAS-CONF-2018-023}.
 
The events are classified into two exclusive categories: at least three leptons (referred to as 3$\ell$ selection, aimed at top squark decays involving $Z$ bosons), or exactly one lepton (referred to as 1$\ell$ selection, aimed at top squark decays involving Higgs bosons). The selection requirements for each of these categories are described below.
 
\subsection{3$\ell$ selection}
 
In this selection, events are accepted if they satisfy a trigger requiring either two electrons, two muons or an electron and a muon.
The requirements imposed offline on the \pt, identification and isolation of the leptons involved in the trigger decision are tighter than those applied online,
so as to be on the trigger efficiency plateau~\cite{TRIG-2016-01}.
The presence of at least three signal leptons (electrons or muons, referred to collectively with the symbol $\ell$), with at least one SF-OS lepton pair whose invariant mass is compatible with the $Z$ boson mass ($|m_{\ell\ell}-m_Z|<15$~\GeV, with $m_Z=91.2$~ \GeV) is required.
In addition, the leading (highest $\pt$) lepton is required to have $\pt>40$~\GeV\ and the subleading to have $\pt>20$~\GeV.
The SF-OS requirements are not applied for the validation of the background induced by fake and non-prompt leptons in Section~\ref{sec:bkgF}.
 
Four overlapping signal regions (SRs) are optimised for the best discovery sensitivity, two for each of the simplified models described in Section~\ref{sec:intro}.
The requirements in each SR are summarised in Table~\ref{tab:SRdefZ}.
 
Signal region \SRH\ is optimised for large $\ninotwo - \ninoone$ mass splittings in the $\stopone\to t\ninotwo$ with $\ninotwo\to h/Z\ninoone$ model. It includes requirements on \mTtwoZ, a variation of the stransverse mass $m_{\textrm{T2}}$ which is used to bind the masses of a pair of particles that are presumed to have each decayed semi-invisibly into one visible and one invisible particle~\cite{mt2,mt22}. In the case of \mTtwoZ, the two visible legs of the two semi-invisible decays are set to be the third leading lepton and the system of the SF-OS lepton pair with an invariant mass closest to $m_Z$.
Models with small mass differences between the $\ninotwo$ and the $\ninoone$ are targeted with \SRL, which instead imposes requirements on the transverse momentum of the SF-OS lepton pair ($p_{\textrm T}^{\ell\ell}$).
 
Two SRs are optimised for the $\stoptwo\to Z\stopone$ with $\stopone \to b ff'\ninoone$ model, \SRSC\ and \SRSB, targeting small and large mass splittings between the $\stoptwo$ and the $\ninoone$, respectively. Due to the overall soft kinematics of the particles in compressed $\stoptwo$ signals, \SRSC\ features upper bounds on the $\pt$ of the third leading lepton and on $p_{\textrm T}^{\ell\ell}$, as well as no requirement on the number of $b$-tagged jets since they are likely to be soft in this scenario. \SRSB\ also includes an upper bound on the third leading lepton $\pt$ but requires the presence of $b$-tagged jets  and large $\met$ and $p_{\textrm T}^{\ell\ell}$.
 
The requirement on the number of signal leptons makes the \SRH\ and \SRL\ selections insensitive to potential contributions from alternative \stopone\ decays with each $\stopone \rightarrow t\ninoone$. The acceptance for mixed decay scenarios, with $\stopone \stopone \rightarrow t \ninoone t \ninotwo$ was found to depend linearly on the branching fraction of \stopone\ to $t\ninotwo$. The signal lepton multiplicity requirement depletes the contribution from the direct production of \stopone\ pairs to \SRSC\ and \SRSB.
 
\begin{table}[!htb]
\caption{\label{tab:SRdefZ} Definition of the signal regions used in the $3\ell$ selection (see text for further description).}
\def\arraystretch{1.2}
\begin{center}
\begin{tabular}{l r r r r}
\noalign{\smallskip}\hline\noalign{\smallskip}
Requirement / Region                            & \SRH       & \SRL       & \SRSC      & \SRSB \\
\noalign{\smallskip}\hline\noalign{\smallskip}
Number of signal leptons                & \multicolumn{4}{c}{$\geq 3$}  \\
Number of SF-OS pairs                   & \multicolumn{4}{c}{$\geq 1$}  \\
Leading lepton $\pt$ [\GeV]             & \multicolumn{4}{c}{$>40$}  \\
Subleading lepton $\pt$ [\GeV]      & \multicolumn{4}{c}{$>20$}  \\
$|m^{\text{SF-OS}}_{\ell\ell}-m_Z|$ [\GeV]     & \multicolumn{4}{c}{$<15$}  \\\hline
Third leading lepton $\pt$ [\GeV]               & $>20$      & $>20$      & $<20$      & $<60$  \\
$n_{\textrm{jets}}$ ($\pt>30$~\GeV)             & $\geq 4$   & $\geq 5$   & $\geq 3$   & $\geq 3$ \\
$n_{b{\textrm{-tagged\,jets}}}$ ($\pt>30$~\GeV) & $\geq 1$   & $\geq 1$   & --         & $\geq 1$ \\
Leading jet $p_\text{T}$ [\GeV]                 & --         & --         & $>150$     & --   \\
Leading $b$-tagged jet $p_\text{T}$ [\GeV]      & --         & $>100$     & --         & --   \\
\MET [\GeV]                                     & $>250$     & $>150$     & $>200$     & $>350$  \\
$p_{\textrm T}^{\ell\ell}$ [\GeV]               & --         & $>150$     & $<50$      & $>150$   \\
\mTtwoZ\ [\GeV]                                   & $>100$     & --         & --         & --     \\
\hline
\end{tabular}
 
\end{center}
\end{table}

\subsection{1$\ell$ selection}
 
Events in this selection are accepted if they satisfy a combination of single-lepton and $\met$ trigger requirements, the latter being used only for events with $\met>230$~\GeV\ and lepton $\pt <30$~\GeV.
The offline requirements on the $\met$, \pt, identification and isolation of the lepton are tighter than those applied online,
so as to be on the trigger efficiency plateau~\cite{TRIG-2016-01}.
The presence of exactly one signal lepton (electron or muon) is required.
 
The identification of Higgs boson candidates decaying into $b$-quarks is performed using a neural network that uses as input the four-momentum and $b$-tagging information of pairs of jets.
The scaled jet $\pt/m_{jj}$ observable is used to
prevent the dijet invariant mass being used as the primary discriminating variable.
The network is trained with the PyTorch package~\cite{NEURIPS2019_9015} using as signal dijet pairs originating from Higgs boson decays in the simulated $\ttbar H$ process, and as background dijet pairs in $\ttbar$ and $\ttbar H$ not originating from Higgs boson decays.
A jet pair is tagged as a Higgs boson candidate if the corresponding neural network score is above a threshold optimised to target models predicting sizeable branching fractions of $\ninotwo$ into Higgs bosons.
The efficiency to correctly identify jet pairs originating from Higgs boson decays in signal events is 50\%--54\% depending on the $\stopone$ mass, as evaluated on a simulated event sample selected by requiring one signal lepton and at least two $b$-tagged jets. Using the same selection, the probability of tagging a jet pair in $\ttbar$ events is approximately 0.05\%.
 
Two overlapping SRs are optimised for the best discovery sensitivity in the $\stopone \to t\ninotwo$ simplified model
discussed in Section~\ref{sec:intro}, and their requirements are summarised in Table~\ref{tab:SRdefH}. Both require the presence of at least four $b$-tagged jets and at least one Higgs boson candidate ($n_{h\textrm{-cand}}$). The SRs include requirements
on the transverse mass $m_{\textrm T}$ (computed as $m_{\textrm T} = \sqrt{2 p_\textrm{T}^{\ell}\met(1 - \cos \Delta\phi)}$ where $\Delta\phi$ is the azimuthal angle between the missing transverse momentum vector and the lepton), and on the object-based $\met$-significance~\cite{ATLAS-CONF-2018-038} ($\cal S$), which is used to discriminate events where the $\met$ is due to invisible particles in the final state from events where the $\met$ is due to poorly measured particles and jets. The requirements on $n_{b{\textrm{-tagged\,jets}}}$ and $n_{h\textrm{-cand}}$ reduce the potential contributions from alternative \stopone\ decays with each $\stopone \rightarrow t\ninoone$ to a negligible level. Analogously to the 3$\ell$ selection, the acceptance for mixed decay scenarios with $\stopone\stopone \rightarrow t\ninoone t\ninotwo$ has been found to depend linearly on the branching fraction of \stopone\ to $t\ninotwo$.
 
Signal region \SRbH\  is optimised for small $\ninotwo - \ninoone$ mass splittings, while large $\ninotwo - \ninoone$ mass splittings are  targeted by \SRbL.

\begin{table}[!htb]
\caption{\label{tab:SRdefH} Definition of the signal regions used in the $1\ell$ selection (see text for further description).}
\def\arraystretch{1.2}
\begin{center}
\begin{tabular}{l r r}
\noalign{\smallskip}\hline\noalign{\smallskip}
Requirement / Region                            & \SRbH       & \SRbL       \\
\noalign{\smallskip}\hline\noalign{\smallskip}
Number of signal leptons                & \multicolumn{2}{c}{$	1$}  \\
$n_{h\textrm{-cand}}$             & \multicolumn{2}{c}{$\geq 1$}  \\
$n_{b{\textrm{-tagged\,jets}}}$ ($\pt>30$~\GeV) &  \multicolumn{2}{c}{$\geq 4$}    \\
$n_{\textrm{jets}}$ ($\pt>60$~\GeV)              & $\geq 4$  & $\geq 6$  \\
\mT\ [\GeV]                                   & $>150$     & $>150$             \\
$\cal S$                                         & $>12$     & $>7$  \\
\hline
\end{tabular}
\end{center}
\end{table}
 
\FloatBarrier
 
\section{Background estimation}
\label{sec:bkg}
 
The dominant SM background contribution to the $3\ell$ SRs is expected to originate from $\ttbar Z$ production, with minor contributions from multi-boson production (mainly $WZ$) and backgrounds containing hadrons misidentified as leptons (hereafter referred to as `fake' leptons) or non-prompt leptons from decays of hadrons (mainly in $t\bar{t}$ events). The main background affecting the $1\ell$ SRs is expected to originate from $\ttbar$ production in association with heavy-flavour quarks.
 
The background from fake/non-prompt (FNP) leptons is estimated in a data-driven way, while the normalisation of the main backgrounds ($\ttbar Z$ and multi-boson in the $3\ell$ selection; $\ttbar$ in the $1\ell$ selection) is obtained by fitting the yield from MC simulation to the observed data in dedicated control regions (CRs) enhanced in a particular background component, and then extrapolating this yield to the SRs.
Backgrounds from other sources, which provide a subdominant contribution to the SRs, are determined from MC simulation only.
 
The expected SM background is determined separately in each SR from a profile likelihood fit~\cite{Cowan:2010js} implemented in the HistFitter framework~\cite{Baak:2014wma}.
The fit uses as a constraint the observed event yield in the fitted regions to adjust the normalisation of the main backgrounds. The quality of the resulting  background model is judged by performing a `background-only' fit to data using exclusively the event yield in the CRs to adjust the normalisation of the backgrounds. The agreement of the resulting background model is compared with the data yields in dedicated validation regions (VRs). Systematic uncertainties related to the MC modelling affect the expected yields in the SRs and CRs, and are taken into account to determine the uncertainty in the background prediction. Each source of uncertainty is described by a single nuisance parameter, and correlations between background processes and selections are taken into account. The VRs, used to assess the quality of the background model, are not used to constrain the nuisance parameters in the fit. The fit does not significantly affect either the uncertainty or the central value of these nuisance parameters. The systematic uncertainties considered in the fit are described in Section~\ref{sec:syst}.
 
\subsection{Fake/non-prompt lepton background}
\label{sec:bkgF}
 
A method similar to that described in Refs.~\cite{top1,top2} is used for the estimation of the FNP lepton background.
Two types of lepton identification criteria are defined for this evaluation: `tight' and `loose', corresponding to the signal and candidate electrons and muons described in Section~\ref{sec:selection}.
With the number of observed events with tight or loose leptons, 
the method estimates the number of events containing prompt or FNP leptons using as input the probability for loose prompt or FNP leptons to satisfy the tight criteria.
The probability for prompt loose leptons to satisfy the tight selection is determined from $\ttbar Z$ MC simulation, applying correction factors obtained by comparing $Z\to\ell^+ \ell^-$ ($\ell = e, \mu$) events in data and MC simulation.
The equivalent probability for loose FNP leptons to satisfy the tight selection is measured in a data sample enhanced in $\ttbar$ using events with one electron and one muon with the same charge plus at least one $b$-tagged jet.
 
The estimates for the FNP background are validated in dedicated regions with selection criteria similar to those defining the 3$\ell$ SRs but with reduced contributions from processes with three prompt leptons.
Two VRs are defined as detailed in Table~\ref{tab:VRF}, with \VRFone\ probing the lepton $\pt$ regime in \SRH\ and \SRL, while \VRFtwo\ includes soft lepton requirements as in \SRSC\ and \SRSB.
To enhance fakes and be mutually exclusive with the SR, events in these VRs are required to have no SF-OS dilepton pairs
and to have at least one different-flavour opposite-sign (DF-OS) dilepton pair. The observed and expected yields in these VRs are shown in Table~\ref{tab:VRF_res}, with a purity of FNP leptons above 55\% and with good agreement between data and the background estimates.
 
The contribution of the FNP background in the 1$\ell$ selection is negligible.
 
\begin{table}[!htb]
\caption{\label{tab:VRF} Definition of the validation regions used for the FNP lepton estimation (see text for further description).}
\def\arraystretch{1.2}
\begin{center}
\begin{tabular}{l r r }
\noalign{\smallskip}\hline\noalign{\smallskip}
Requirement / Region                   & \VRFone & \VRFtwo \\
\noalign{\smallskip}\hline\noalign{\smallskip}
Number of signal leptons                & \multicolumn{2}{c}{$\geq 3$}  \\
Number of SF-OS pairs                   & \multicolumn{2}{c}{$0$}  \\
Number of DF-OS pairs                   & \multicolumn{2}{c}{$\geq 1$}  \\
Leading lepton $\pt$ [\GeV]             & \multicolumn{2}{c}{$>40$}  \\
Subleading lepton $\pt$ [\GeV]      & \multicolumn{2}{c}{$>20$}  \\\hline
Third leading lepton $\pt$ [\GeV]       & $>20$		& $<60$	 \\
$n_{\textrm{jets}}$ ($\pt>30$~\GeV)     & $\geq 3$& $\geq 3$ \\
$n_{b{\textrm{-tagged\,jets}}}$ ($\pt>30$~\GeV)   & $\geq 1$ & --   \\
\MET [\GeV]                             & $>50$   & $>150$   \\
\hline
\end{tabular}
\end{center}
\end{table}

\begin{table}[!tbh]
\caption{Background fit results for the FNP validation regions.
The `Others' category is dominated by $\ttbar W$ production and also contains the contributions from $\ttbar h$, $\ttbar WW$, $\ttbar t$, $\ttbar \ttbar$, $Wh$, and $Zh$ production.
Combined statistical and systematic uncertainties are given.
Some of the uncertainties are correlated and therefore the overall sum in quadrature does not necessarily add to the total systematic uncertainty.
The number of $\ttZ$ and multi-boson background events is estimated as described in Section~\ref{sec:BkgPrompt}.
\label{tab:VRF_res}
}
\begin{center}
\setlength{\tabcolsep}{0.0pc}
{\small
\begin{tabular*}{0.5 \textwidth}{@{\extracolsep{\fill}}lc rcl c rcl}
\noalign{\smallskip}\hline\noalign{\smallskip}
&&  \multicolumn{3}{c}{\VRFone}             &&  \multicolumn{3}{c}{\VRFtwo}              \\[-0.05cm]
\noalign{\smallskip}\hline\noalign{\smallskip}
Observed events          && \multicolumn{3}{c}{$84$}            && \multicolumn{3}{c}{$98$}                  \\
\noalign{\smallskip}\hline\noalign{\smallskip}
Total (post-fit) SM events         && $104$ & $\pm$ & $28$        && $98$ & $\pm$ & $33$             \\
\noalign{\smallskip}\hline\noalign{\smallskip}
Post-fit, multi-boson    && $0.7$ & $\pm$ & $0.2$          && $2.7$ & $\pm$ & $0.7$            \\
Post-fit, \ttZ           && $9.1$ & $\pm$ & $1.7$          && $2.6$ & $\pm$ & $0.6$          \\
Fake/non-prompt leptons    && $54$ & $\pm$ & $27$        && $76$ & $\pm$ & $33$            \\
$tZ$, $tWZ$                && $0.9$ & $\pm$ & $0.5$          && $0.40$ & $\pm$ & $0.21$         \\
Others                     && $39$ & $\pm$ & $6$         && $16.2$ & $\pm$ & $3.1$           \\
\noalign{\smallskip}\hline\noalign{\smallskip}
\end{tabular*}
}
\end{center}
\end{table}
 
\subsection{Estimation of the $\ttbar Z$ and multi-boson background in the 3$\ell$ selection}
\label{sec:BkgPrompt}
 
The two dedicated control regions used for the $\ttbar Z$ (\CRTZ) and multi-boson (\CRVV) background estimation in the 3$\ell$ selection are defined in Table~\ref{tab:CRdef}.
To ensure mutual exclusion with the SRs, only events with $50~\GeV\ <\MET <100$~\GeV\ are included in \CRTZ, while a $b$-tagged jet veto and a $50~\GeV\ <\MET <200$~\GeV\ requirement are applied in \CRVV.
 
To validate the background estimates and provide a statistically independent cross-check of the extrapolation to the SRs, three validation regions are defined, as shown in Table~\ref{tab:CRdef}. The \VRTZ\ region primarily validates the \ttZ\ background estimate. The \VRVV\ and \VRVVb\ regions validate the multi-boson background estimate, the former focusing on the extrapolation in jet multiplicity and the latter releasing the $b$-tagged jet veto. The overlap between these multi-boson VRs is around 50\%.
 
Table~\ref{tab:CRVRZ} shows the observed and expected yields in the CRs and VRs for each background source, and Figure~\ref{fig:var_CR} shows the jet multiplicity distribution after the background fit for these CRs and VRs.
The normalisation factors for the $\ttbar Z$ and multi-boson backgrounds do not differ from unity by more than 20\%
and the post-fit MC-simulated jet multiplicity distributions agree well with the data.
 
\begin{table}[!htb]
\caption{\label{tab:CRdef} Definition of the control and validation regions used for the $\ttbar Z$ and multi-boson background estimation.}
 
\begin{center}
\def\arraystretch{1.2}
\begin{tabular}{lrrrrr}
\noalign{\smallskip}\hline\noalign{\smallskip}
Requirement / Region                            & \CRTZ    & \VRTZ    & \CRVV    & \VRVV    & \VRVVb \\
\hline
Number of signal leptons                & \multicolumn{4}{c}{$\geq 3$}  \\
Number of SF-OS pairs                   & \multicolumn{4}{c}{$\geq 1$}  \\
Leading lepton $\pt$ [\GeV]             & \multicolumn{4}{c}{$>40$}  \\
$|m^{\text{SF-OS}}_{\ell\ell}-m_Z|$ [\GeV]     & \multicolumn{4}{c}{$<15$}  \\\hline
Second leading lepton $p_\text{T}$ [\GeV]                & $>20$    & $>20$    & $>20$    & $>40$    & $>40$   \\
Third leading lepton $p_\text{T}$ [\GeV]                 & $>20$    & $>20$    & $>20$    & $>40$    & $>40$   \\
$n_{\textrm{jets}} (\pT>30$~\GeV)                & $\geq 4$ & $\geq 3$ & $\geq 3$ & $\geq 3$ & $3$ \\
$n_{b{\textrm{-tagged\,jets}}}$ ($\pt>30$~\GeV) & $\geq 1$ & $\geq 1$ & 0        & $ 0$     & $\geq 0$ \\
\MET [\GeV]                                      & 50--100  & 100--150 & 50--200  & 200--300 & 200--300 \\
\hline
\end{tabular}
\end{center}
\end{table}
 
\begin{figure}[htb!]
\begin{center}
\begin{subfigure}[t]{0.45\textwidth}
\centering
\includegraphics[width=\textwidth]{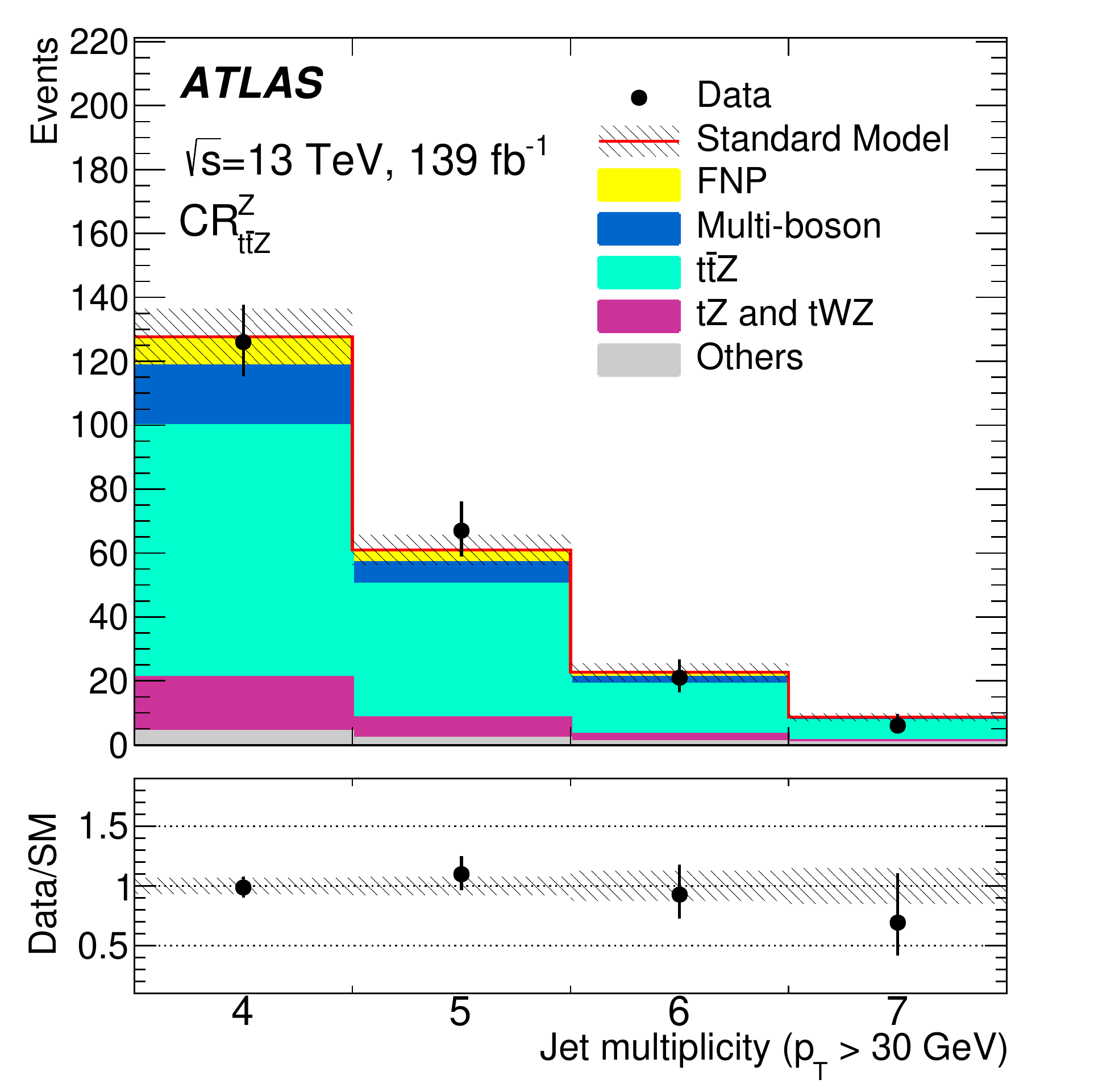}
\caption{}
\end{subfigure}
\begin{subfigure}[t]{0.45\textwidth}
\centering
\includegraphics[width=\textwidth]{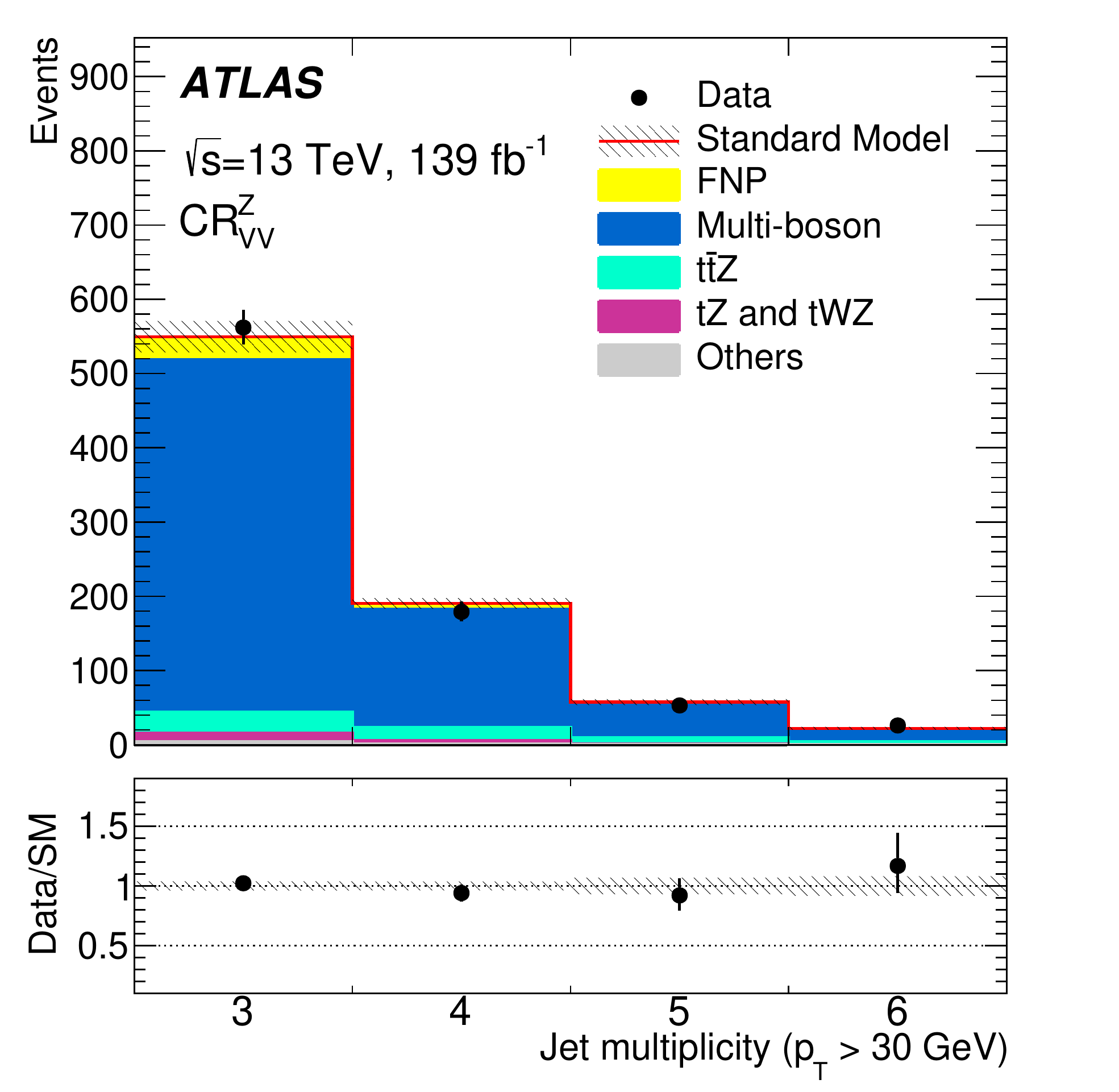}
\caption{}
\end{subfigure}
 
\begin{subfigure}[t]{0.45\textwidth}
\centering
\includegraphics[width=\textwidth]{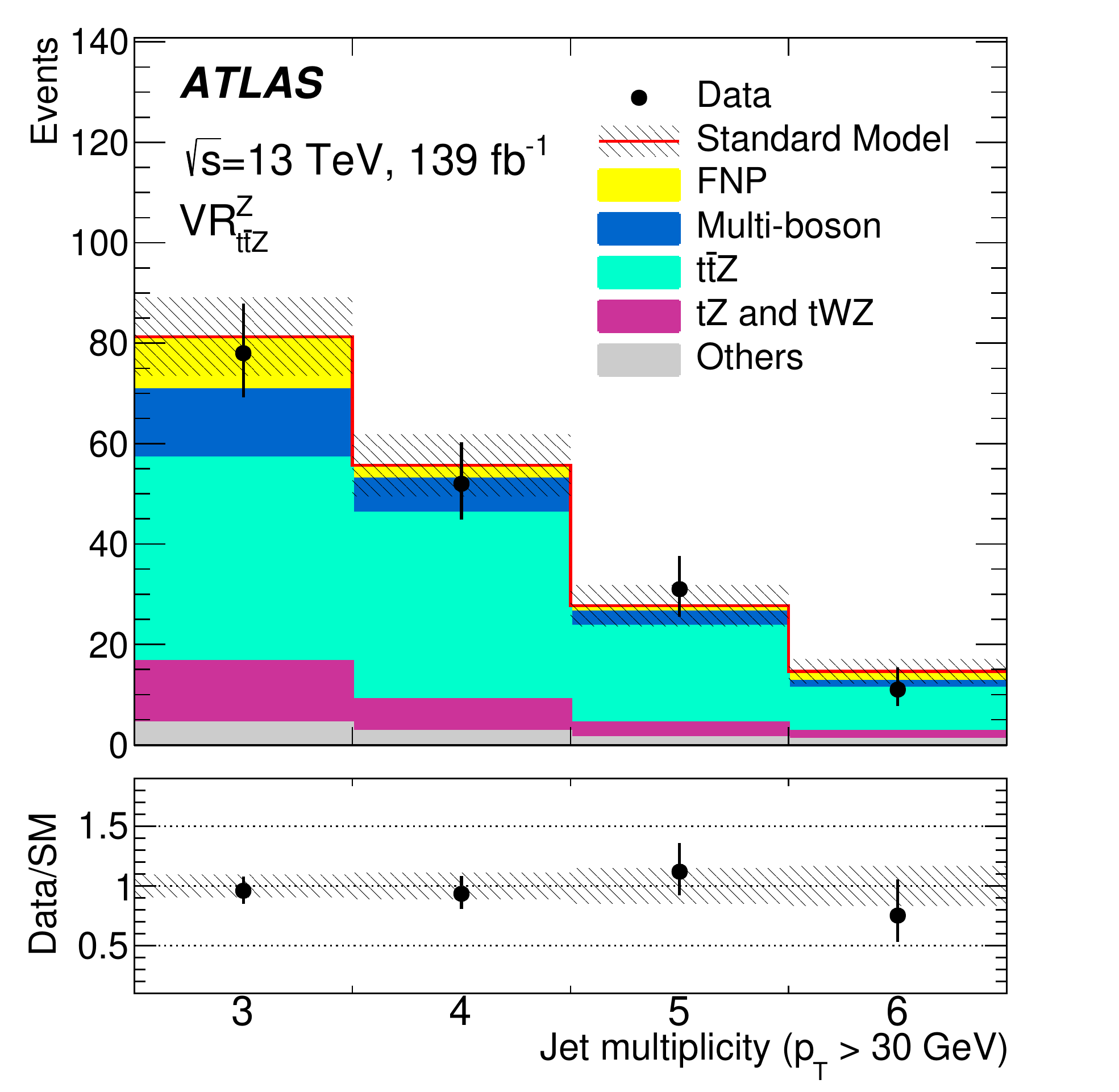}
\caption{}
\end{subfigure}
\begin{subfigure}[t]{0.45\textwidth}
\centering
\includegraphics[width=\textwidth]{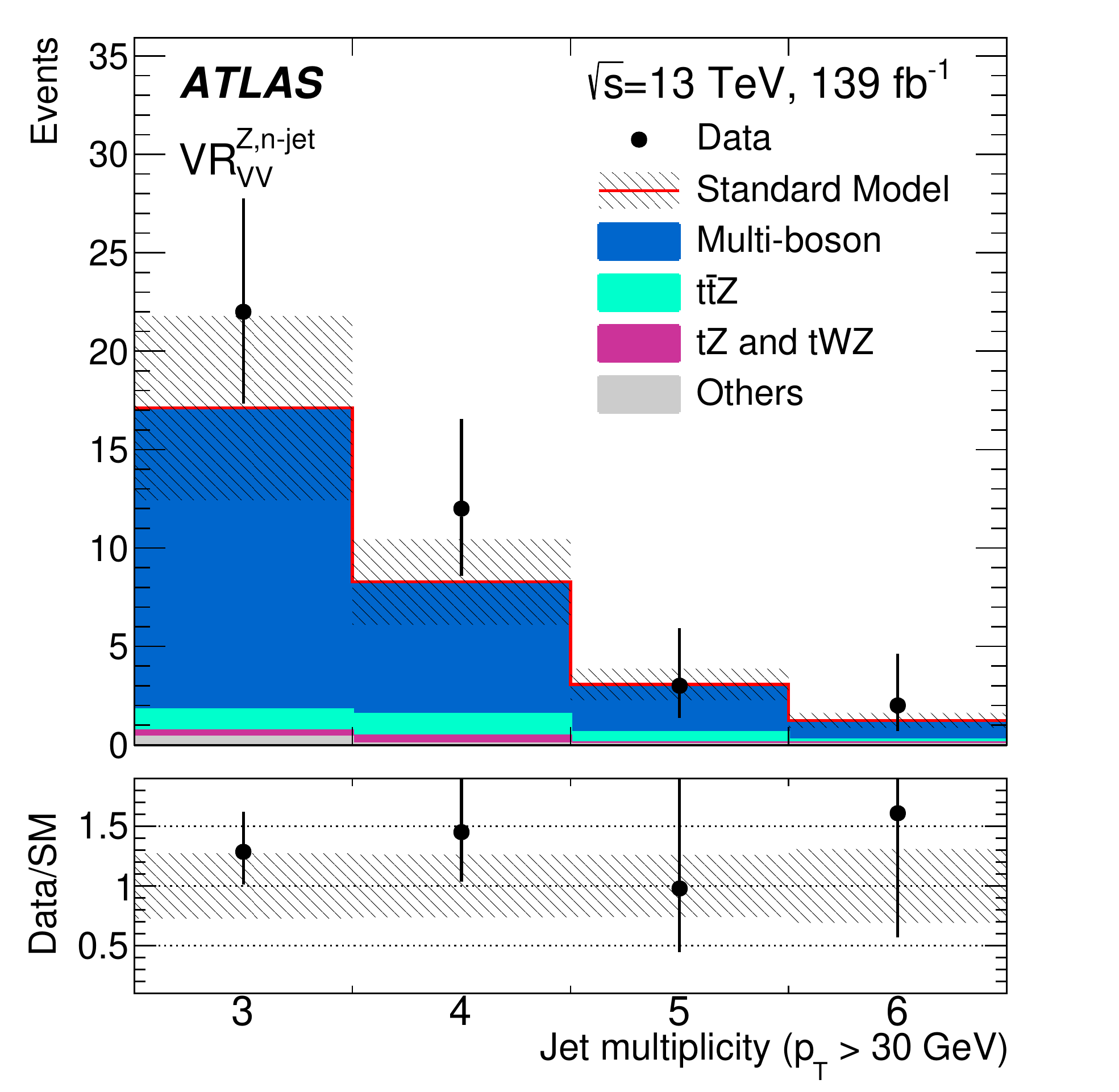}
\caption{}
\end{subfigure}
\end{center}
\caption{Jet multiplicity distributions in control and validation regions (a) \CRTZ, (b) \CRVV, (c) \VRTZ\ and (d) \VRVV after normalising the $\ttZ$ and multi-boson background processes via the simultaneous fit described in Section~\ref{sec:bkg}. The contributions from all SM backgrounds are shown as a histogram stack; the bands represent the total uncertainty in the background prediction.
The `Others' category contains the contributions from $\ttbar h$, $\ttbar W$, $\ttbar WW$, $\ttbar t$, $\ttbar \ttbar$, $Wh$, and $Zh$ production.
The `FNP' category represents the background from fake or non-prompt leptons.
The last bin in each figure contains the overflow.
The lower panels show the ratio of the observed data to the total SM background prediction, with bands representing the total uncertainty in the background prediction.}
\label{fig:var_CR}
\end{figure}
 
\begin{table}[!tbh]
\caption{Background fit results for the control and validation regions for the $\ttbar Z$ and multi-boson backgrounds. The pre-fit predictions from MC simulation are given for comparison for those backgrounds ($\ttZ$, multi-boson) that are normalised to data.
The `Others' category contains the contributions from $\ttbar h$, $\ttbar W$, $\ttbar WW$, $\ttbar t$, $\ttbar \ttbar$, $Wh$, and $Zh$ production.
Combined statistical and systematic uncertainties are given.
Some of the uncertainties are correlated and therefore the overall sum in quadrature does not necessarily add to the total systematic uncertainty.
The number of events with fake/non-prompt leptons is estimated with the data-driven technique described in Section~\ref{sec:bkgF}.
\label{tab:CRVRZ}
}
\begin{center}
\setlength{\tabcolsep}{0.0pc}
{\small
\begin{tabular*}{\textwidth}{@{\extracolsep{\fill}}lc rcl c rcl c rcl c rcl c rcl }
\noalign{\smallskip}\hline\noalign{\smallskip}
&& \multicolumn{3}{c}{\CRTZ}    && \multicolumn{3}{c}{\VRTZ}       && \multicolumn{3}{c}{\CRVV}        && \multicolumn{3}{c}{\VRVV}   && \multicolumn{3}{c}{\VRVVb}       \\[-0.05cm]
\noalign{\smallskip}\hline\noalign{\smallskip}
Observed events          && \multicolumn{3}{c}{$220$}       && \multicolumn{3}{c}{$172$}           && \multicolumn{3}{c}{$820$}           && \multicolumn{3}{c}{$39$}              && \multicolumn{3}{c}{$34$}            \\
\noalign{\smallskip}\hline\noalign{\smallskip}
Total (post-fit) SM events  && $220$ & $\pm$ & $15$           && $179$ & $\pm$ & $16$   && $820$ & $\pm$ & $29$    && $30$ & $\pm$ & $8$          && $26$ & $\pm$ & $6$      \\
\noalign{\smallskip}\hline\noalign{\smallskip}
Post-fit, multi-boson      && $28$ & $\pm$ & $7$    && $25$ & $\pm$ & $8$     && $698$ & $\pm$ & $34$ && $26$ & $\pm$ & $8$         && $17$ & $\pm$ & $6$       \\
Post-fit, \ttZ             && $142$ & $\pm$ & $22$  && $105$ & $\pm$ & $20$   && $57$ & $\pm$ & $11$  && $2.8$ & $\pm$ & $0.6$          && $5.4$ & $\pm$ & $1.2$        \\
Fake/non-prompt leptons      && $15.1$ & $\pm$ & $1.7$    && $16.7$ & $\pm$ & $1.9$     && $41$ & $\pm$ & $15$  && \multicolumn{3}{c}{$<$1.5}          && \multicolumn{3}{c}{$0.9_{-0.9}^{+1.1}$} \\
$tZ$, $tWZ$                  && $27$ & $\pm$ & $14$   && $23$ & $\pm$ & $12$    && $19$ & $\pm$ & $10$   && $0.9$ & $\pm$ & $0.5$          && $1.7$ & $\pm$ & $0.9$        \\
Others                       && $8.0$ & $\pm$ & $1.4$     && $9.7$ & $\pm$ & $2.1$      && $5.8$ & $\pm$ & $1.3$    && $0.46$ & $\pm$ & $0.09$          && $0.62$ & $\pm$ & $0.12$         \\
\noalign{\smallskip}\hline\noalign{\smallskip}
Pre-fit, multi-boson      && $35.4$ & $\pm$ & $3.5$    && $31$ & $\pm$ & $7$    && $870$ & $\pm$ & $200$  && $32$ & $\pm$ & $7$          && $22$ & $\pm$ & $5$  \\
Pre-fit, \ttZ              && $154$ & $\pm$ & $14$  && $114$ & $\pm$ & $5$   && $61.7$ & $\pm$ & $3.4$  && $3.1$ & $\pm$ & $0.4$          && $5.8$ & $\pm$ & $0.6$    \\
\noalign{\smallskip}\hline\noalign{\smallskip}
\end{tabular*}
}
\end{center}
\end{table}
 
\FloatBarrier
\newpage

\subsection{Estimation of the $\ttbar$ background in the 1$\ell$ selection}
\label{sec:Bkg1L}
 
The $\ttbar$ background represents more than 70\% of the total background in the 1$\ell$ SRs and it is estimated with the aid of a dedicated control region (\CRTT), defined in Table~\ref{tab:CRdefH}.
Compared to the SRs in Table~\ref{tab:SRdefH}, this region does not apply any selection on the number of Higgs boson candidates and features relaxed requirements on \mT, while
only events with exactly four jets and $7<{\cal S}<10$ are included in this CR to ensure orthogonality with the SRs.
Table~\ref{tab:CRVRH} shows the observed and expected yields in the CR for each background source, and Figure~\ref{fig:var_CRH} shows the distribution of the  transverse mass  and number of Higgs boson candidtes after the background fit.
The normalisation factor for the $\ttbar$ background was found to be $1.09 \pm 0.13$.
 
The extrapolation across $\cal S$ and \mT\ between \CRTT\ and the SRs is tested in several validation regions.
Figure~\ref{fig:VRH} shows the definition, the observed number of events and expected yields in these regions, which feature either the same $b{\textrm{-tagged\,jets}}$ multiplicity as the SRs with relaxed \mT\ requirements, or the same \mT\ selections as the SRs requiring exactly three $b$-tagged jets. Good agreement between the background estimation and the data is observed in all these validation regions.
 
\begin{table}[!htb]
\caption{\label{tab:CRdefH} Definition of the control region used for the $\ttbar$ background estimation.}
 
\begin{center}
\def\arraystretch{1.2}
\begin{tabular}{lrHHHH}
\noalign{\smallskip}\hline\noalign{\smallskip}
Requirement / Region                            & \CRTT    & \VRHmT & \VRLmT   &  \VRH &  \VRL \\
\hline
Number of signal leptons                & $1$   \\
$n_{h\textrm{-cand}}$             & - &$\geq 1$ & $\geq 1$ & $\geq 1$ & $\geq 1$ \\\hline
$n_{b{\textrm{-tagged\,jets}}}$ ($\pt>30$~\GeV) &  $\geq 4$   &  $\geq 4$ &  $\geq 4$  & 3 & 3 \\
$n_{\textrm{jets}}$ ($\pt>60$~\GeV)                      & 4                 & $\geq 5$  & $\geq 4$ & $\geq 5$  & $\geq 4$   \\
\mT\ [\GeV]                                                                 & $>100$       & 100--150 & 100--150 & $>150$ & $>150$ \\
$\cal S$                                                     & 7--10      & $>7$         & $>10$   & $>7$         & $>10$  \\
\hline
\end{tabular}
\end{center}
\end{table}
 
\begin{figure}[htb!]
\begin{center}
\begin{subfigure}[t]{0.45\textwidth}
\centering
\includegraphics[width=\textwidth]{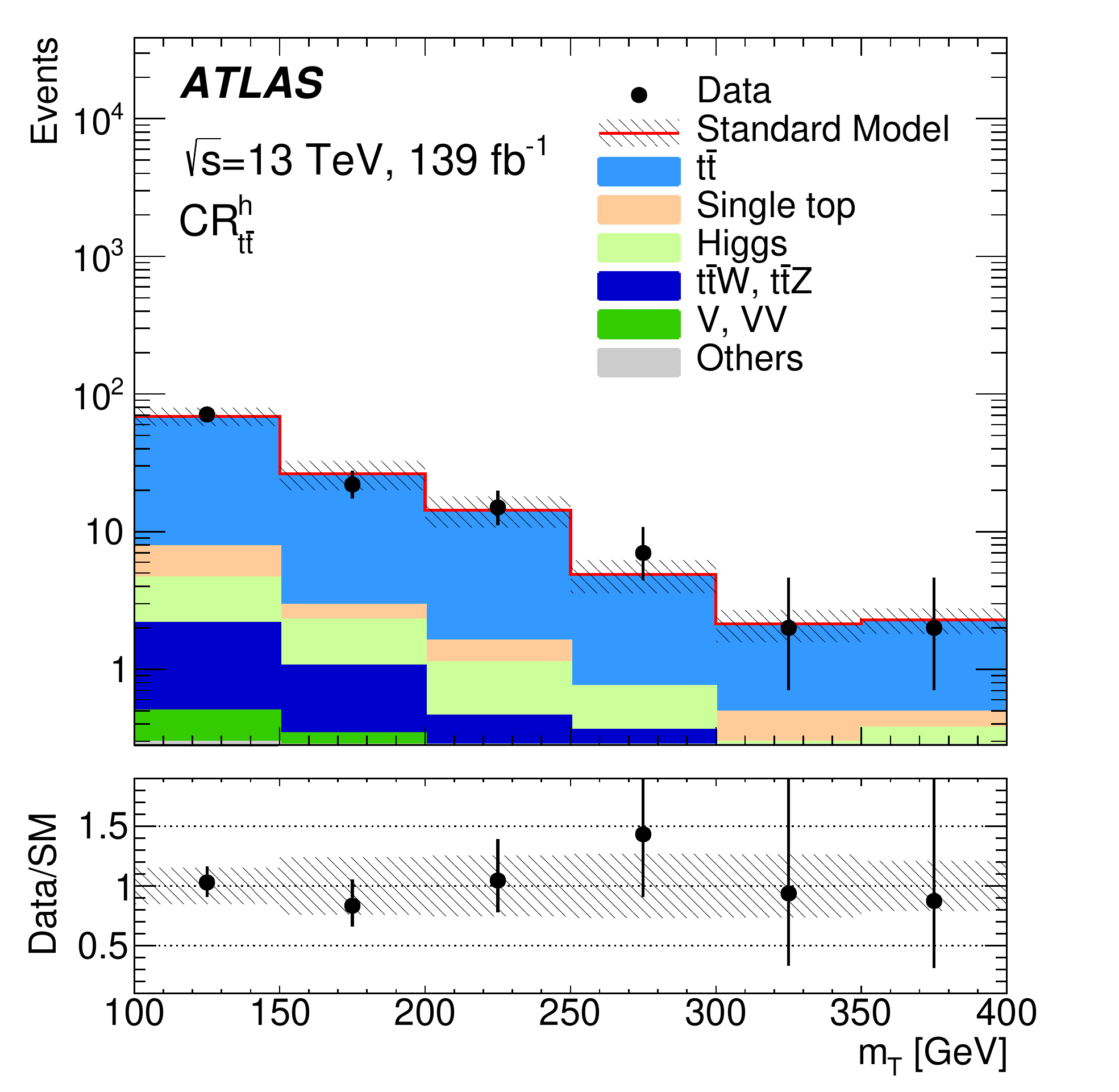}
\caption{}
\end{subfigure}
\begin{subfigure}[t]{0.45\textwidth}
\centering
\includegraphics[width=\textwidth]{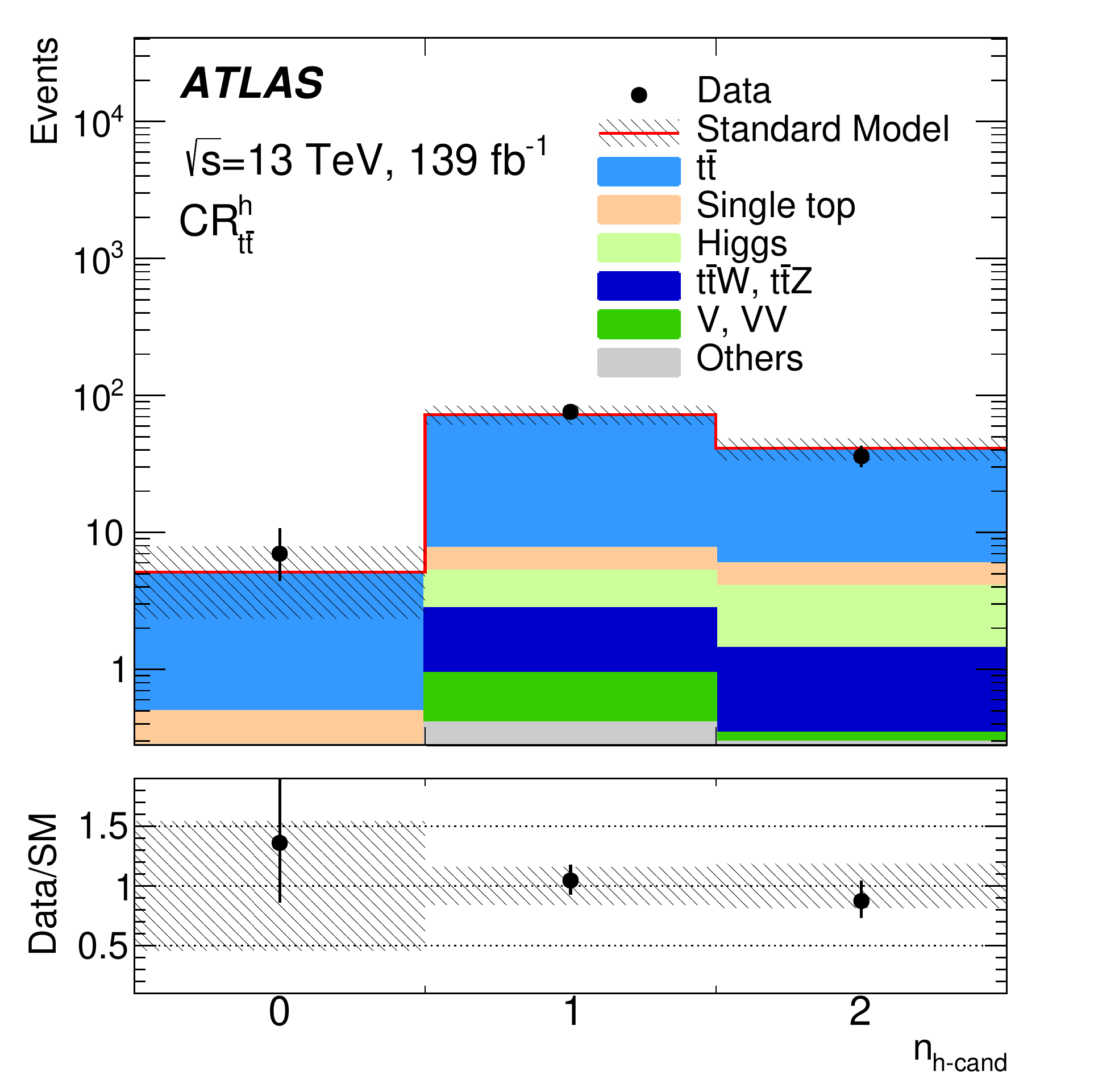}
\caption{}
\end{subfigure}
\end{center}
\caption{
(a) Transverse mass and (b) number of Higgs boson candidates distributions in \CRTT\  after normalising the $\ttbar$ background process via the simultaneous fit described in Section~\ref{sec:bkg}.
The contributions from all SM backgrounds are shown as a histogram stack; the bands represent the total uncertainty in the background prediction.
The `Higgs' category contains the contributions from gluon--gluon fusion, vector-boson fusion, $Wh$, $Zh$ and $\ttbar h$ production.
The `Others' category contains the contributions from $\ttbar WW$, $\ttbar WZ$, $\ttbar t$ and $\ttbar \ttbar$ production.
The last bin in each figure contains the overflow.
The lower panels show the ratio of the observed data to the total SM background prediction, with bands representing the total uncertainty in the background prediction.}
\label{fig:var_CRH}
\end{figure}

\begin{table}[!tbh]
\caption{Background fit results for the $\ttbar$ background control region in the $1\ell$ selection. The pre-fit predictions from MC simulation are given for comparison for those backgrounds ($\ttbar$) that are normalised to data.
The `Higgs' category contains the contributions from gluon--gluon fusion, vector-boson fusion, $Wh$, $Zh$ and $\ttbar h$ production.
The `Others' category contains the contributions from $\ttbar WW$, $\ttbar WZ$, $\ttbar t$ and $\ttbar \ttbar$ production.
Combined statistical and systematic uncertainties are given.
Some of the uncertainties are correlated and therefore the overall sum in quadrature does not necessarily add to the total systematic uncertainty.
\label{tab:CRVRH}
}
\begin{center}
\setlength{\tabcolsep}{0.0pc}
{\small
\begin{tabular*}{0.4\textwidth}{@{\extracolsep{\fill}}l c rcl}
\noalign{\smallskip}\hline\noalign{\smallskip}
&& \multicolumn{3}{c}{\CRTT}     \\[-0.05cm]
\noalign{\smallskip}\hline\noalign{\smallskip}
Observed events          && \multicolumn{3}{c}{$119$}                   \\
\noalign{\smallskip}\hline\noalign{\smallskip}
Total (post-fit) SM events     && $119$ & $\pm$ & $11$              \\
\noalign{\smallskip}\hline\noalign{\smallskip}
Post-fit, $\ttbar$         && $105$ & $\pm$ & $11$              \\
$V$, $VV$         && $0.6$ & $\pm$ & $0.5$              \\
$\ttbar W$, $\ttbar Z$       && $3.0$ & $\pm$ & $0.7$              \\
Higgs       && $5.1$ & $\pm$ & $ 1.8$              \\
Single top         && $4.6$ & $\pm$ & $1.5$              \\
Others        && $0.73$ & $\pm$ & $0.16$              \\
\noalign{\smallskip}\hline\noalign{\smallskip}
Pre-fit, $\ttbar$       && $96$ & $\pm$ & $6$              \\
\noalign{\smallskip}\hline\noalign{\smallskip}
\end{tabular*}
}
\end{center}
\end{table}

\begin{figure}[htb!]
\begin{center}
\includegraphics[width=0.95\textwidth]{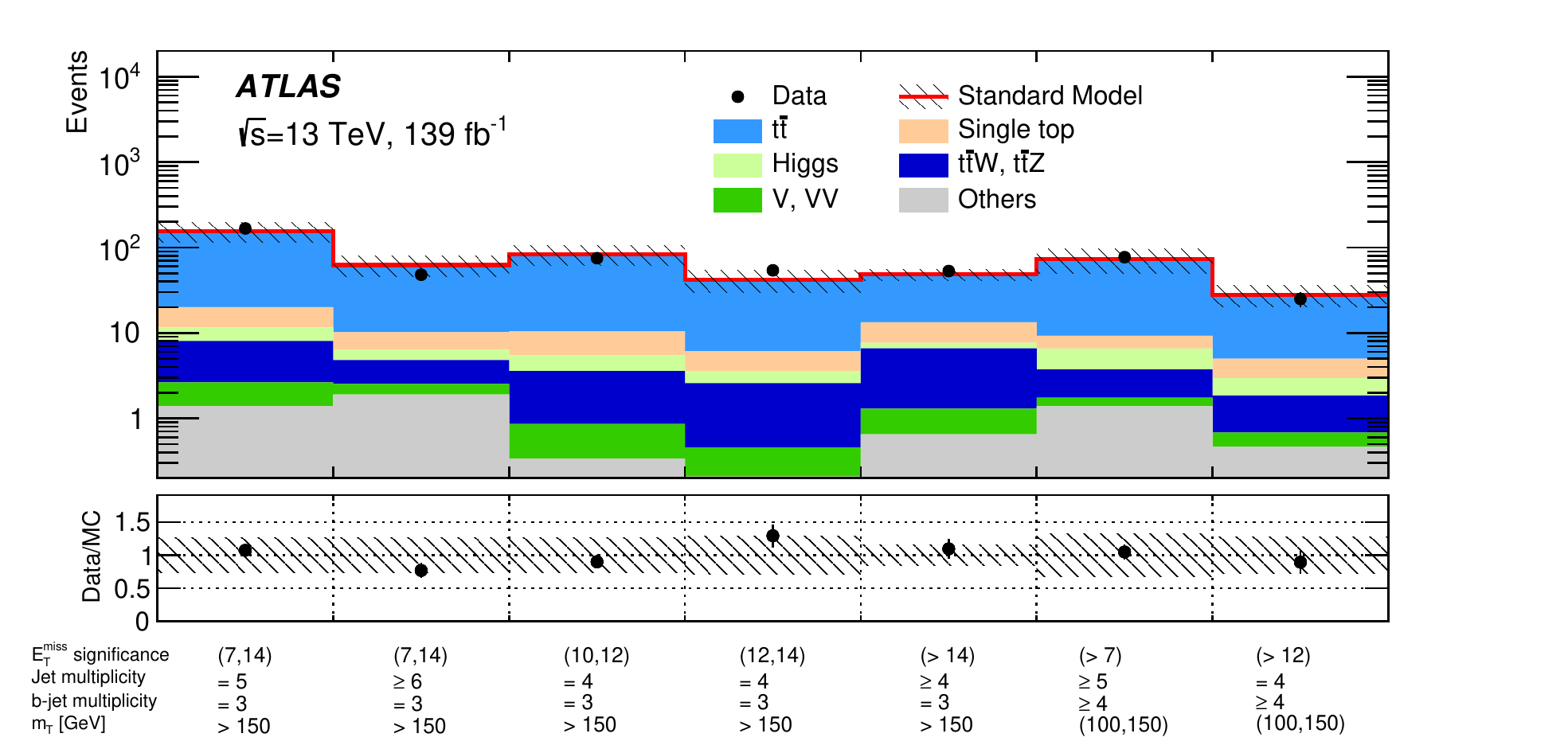}
\caption{Comparison of the observed and expected event yields in the different kinematic regions used to validate the $\ttbar$ background estimation in the 1$\ell$ selection.
The `Higgs' category contains the contributions from gluon--gluon fusion, vector-boson fusion, $Wh$, $Zh$ and $\ttbar h$ production.
The `Others' category contains the contributions from $\ttbar WW$, $\ttbar WZ$, $\ttbar t$ and $\ttbar \ttbar$ production.
The lower panel shows the ratio of the observed data to the total SM background prediction, with bands representing the total uncertainty in the background prediction.}
\label{fig:VRH}
\end{center}
\end{figure}
 
\FloatBarrier
 
\section{Systematic uncertainties}
\label{sec:syst}
 
The main sources of systematic uncertainty affecting the analysis SRs are related to the theoretical and modelling uncertainties in the background, the limited number of events in the CRs and MC simulated samples, the uncertainties in the FNP probabilities, as well as the jet energy scale and resolution. The effects of the systematic uncertainties are evaluated for all signal samples and background processes. Since the normalisation of the dominant background processes is extracted in dedicated CRs, the systematic uncertainties only affect the extrapolation to the SRs in these cases.
Figure~\ref{fig:syst} summarises the contributions from the different sources of systematic uncertainty
to the total SM background predictions in the signal regions.
 
The jet energy scale and resolution uncertainties are derived as a function of the \pT and $\eta$ of the jet, as well as of the pile-up conditions and the jet flavour composition (more like a quark or a gluon) of the selected jet sample. They are determined using a combination of data and simulated event samples, through  measurements of the jet response asymmetry in dijet, $Z$+jet and $\gamma+$jet events~\cite{ATL-PHYS-PUB-2015-015,PERF-2016-04}.
 
Systematic uncertainties in the $b$-tagging efficiency are estimated by varying the $\eta$-, $\pT$- and flavour-dependent scale factors applied to each jet in the simulation within a range that reflects the systematic uncertainty in the measured tagging efficiency and mis-tag rates in data~\cite{FTAG-2018-01, ATLAS-CONF-2018-001,  ATLAS-CONF-2018-006}.
 
Other detector-related systematic uncertainties, such as those related to the \MET modelling, as well as lepton reconstruction efficiency, energy scale and energy resolution are found to have a small impact on the results.
 
\begin{figure}[htb!]
\begin{center}
\includegraphics[width=0.95\textwidth]{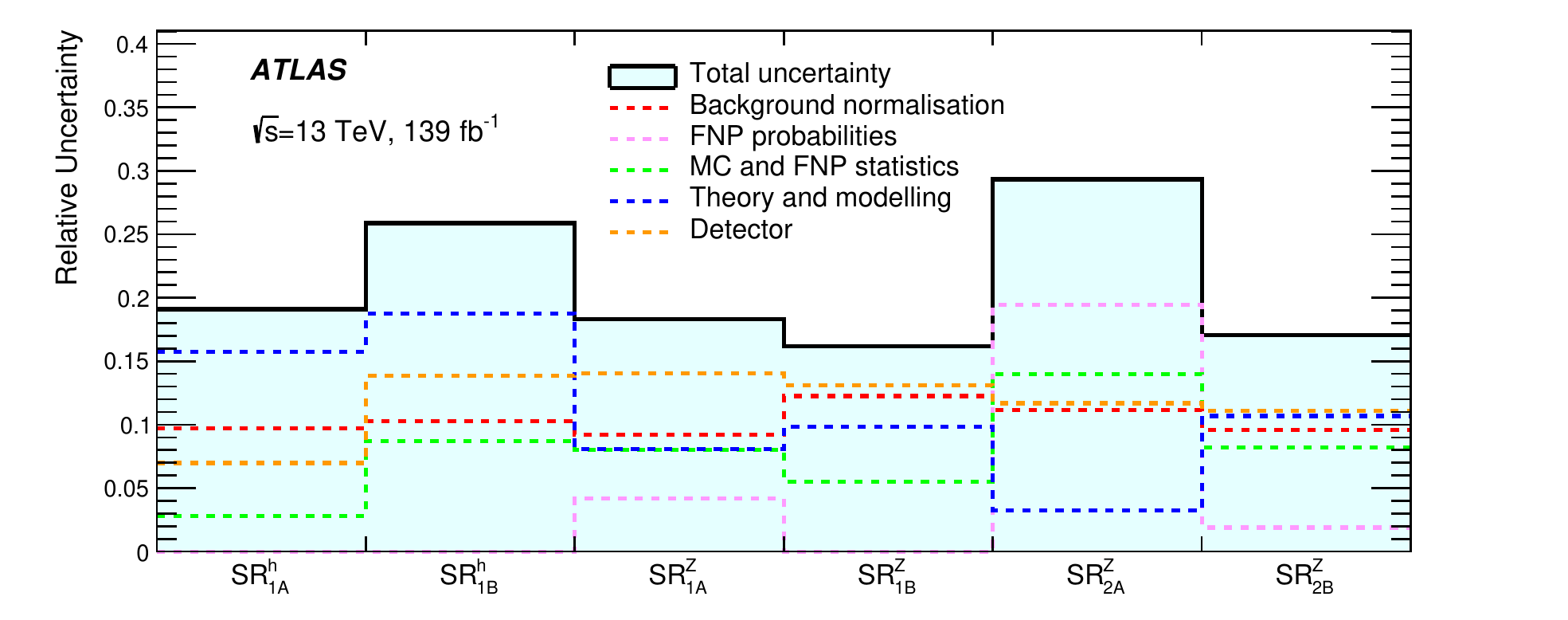}
\end{center}
\caption{Comparison of the relative uncertainty for the total background yield in each SR, including the contribution from the different sources of uncertainty.
The `Detector' category contains all detector-related systematic uncertainties and is dominated by jet energy scale and resolution, and $b$-tagging uncertainties.
}
\label{fig:syst}
\end{figure}
 
Systematic uncertainties are assigned to the FNP background estimation to account for different compositions (heavy flavour, light  flavour or conversions) between the signal and control regions, as well as the contamination from prompt leptons in the regions used to measure the FNP probabilities.
 
The diboson background MC modelling uncertainties are estimated by varying the renormalisation, factorisation and resummation scales used to generate the samples~\cite{ATL-PHYS-PUB-2016-002}.
For the $\ttbar Z$ background, uncertainties due to parton shower and hadronisation modelling are evaluated by comparing the predictions from \AMCATNLO interfaced to \PYTHIA and \HERWIG 7.0.4~\cite{Bellm:2015jjp}, while the uncertainties related to the choice of renormalisation and factorisation scales are assessed by varying the corresponding event generator parameters up and down by a factor of two around their nominal values~\cite{ATL-PHYS-PUB-2016-005}.
For the $\ttbar$ background, uncertainties due to parton shower and hadronisation modelling are evaluated by comparing the predictions from \POWHEGBOX interfaced to \PYTHIA and \HERWIG 7.0.4~\cite{Bellm:2015jjp}, while the uncertainties due the choice of generator are evaluated by comparing the predictions from \POWHEGBOX and \AMCATNLO both interfaced to \PYTHIA.
Variations of the $\ttbar$ initial- and final-state radiation, renormalisation and factorisation scales are also considered~\cite{ATL-PHYS-PUB-2018-009}.
 
The cross-sections used to normalise the MC samples are varied according to the uncertainty in the cross-section calculation, i.e.\ 6\% for diboson, 12\% for $\ttbar Z$, 13\% for $\ttbar W$~\cite{Alwall:2014hca} 
and 5\% for single top production. For $\ttbar WW$, $tZ$, $tWZ$, $\ttbar h$, $Wh$, $Zh$, $\ttbar t$, $\ttbar\ttbar$, and triboson production processes, which constitute a small background, a 50\% uncertainty in the event yields is assumed.

\section{Results}
\label{sec:result}
 
The observed number of events and expected yields are shown in Tables~\ref{tab:results} and~\ref{tab:resultsH} for each of the inclusive 3$\ell$ and 1$\ell$ SRs, respectively. Figures~\ref{fig:kin_SR} and~\ref{fig:kin_SRH} show kinematic distributions after applying all the SR selection requirements except those on $\MET$, $p_{\textrm T}^{\ell\ell}$ or $\cal S$ and jet multiplicity.
The data agree with the SM background predictions and these results are interpreted as exclusion limits for several beyond-the-SM (BSM) scenarios.
 
The HistFitter framework, which utilises a profile-likelihood-ratio test statistic~\cite{Cowan:2010js}, is used to estimate 95\% confidence intervals using the CL$_\mathrm{s}$ prescription~\cite{Read_CLs}.
The likelihood is built from the product of probability density functions describing the observed numbers of events in the SR and the associated CRs. The statistical uncertainties in the CRs and SRs are modelled using Poisson distributions. Systematic uncertainties enter the likelihood as nuisance parameters that are constrained by Gaussian distributions whose widths correspond to the sizes of these uncertainties.
Table~\ref{tab:results} also shows upper limits (at the 95\% CL)  on the number of BSM events $S^{95}$, and on the visible BSM cross-section $\sigma_{\mathrm{vis}} = S_{\textrm{obs}}^{95}/\int\!{\cal L}\textrm{d}t$, 
defined as the product of the production cross-section, acceptance and efficiency.
 
Model-dependent limits are also set in specific classes of SUSY models.
For each signal hypothesis, the background fit is repeated taking into account the signal contamination in the CRs, which is found to be below 12\% for signal models close to the existing exclusion limits~\cite{SUSY-2016-20}.
Correlations of the uncertainties between the SM backgrounds and the signals are taken into account.
 
To enhance the exclusion power in the SUSY models considered, a `shape-fit' approach is used where several mutually exclusive bins in different kinematic variables are defined in order to take advantage of the different signal-to-background ratios in the different bins. Table~\ref{tab:SRexcl} shows the definition of these bins, which loosen a few of the requirements of the discovery SRs to increase the acceptance for different classes of models across the phase space (Tables~\ref{tab:SRdefZ} and~\ref{tab:SRdefH}). \SRSC is used in the exclusion fits with no changes.
The observed number of events and expected yields in all these bins are shown in Figures~\ref{fig:SRsumm} and~\ref{fig:SRHsumm}.
 
\begin{figure}[htb!]
\begin{center}
\begin{subfigure}[t]{0.4\textwidth}
\centering
\includegraphics[width=\textwidth]{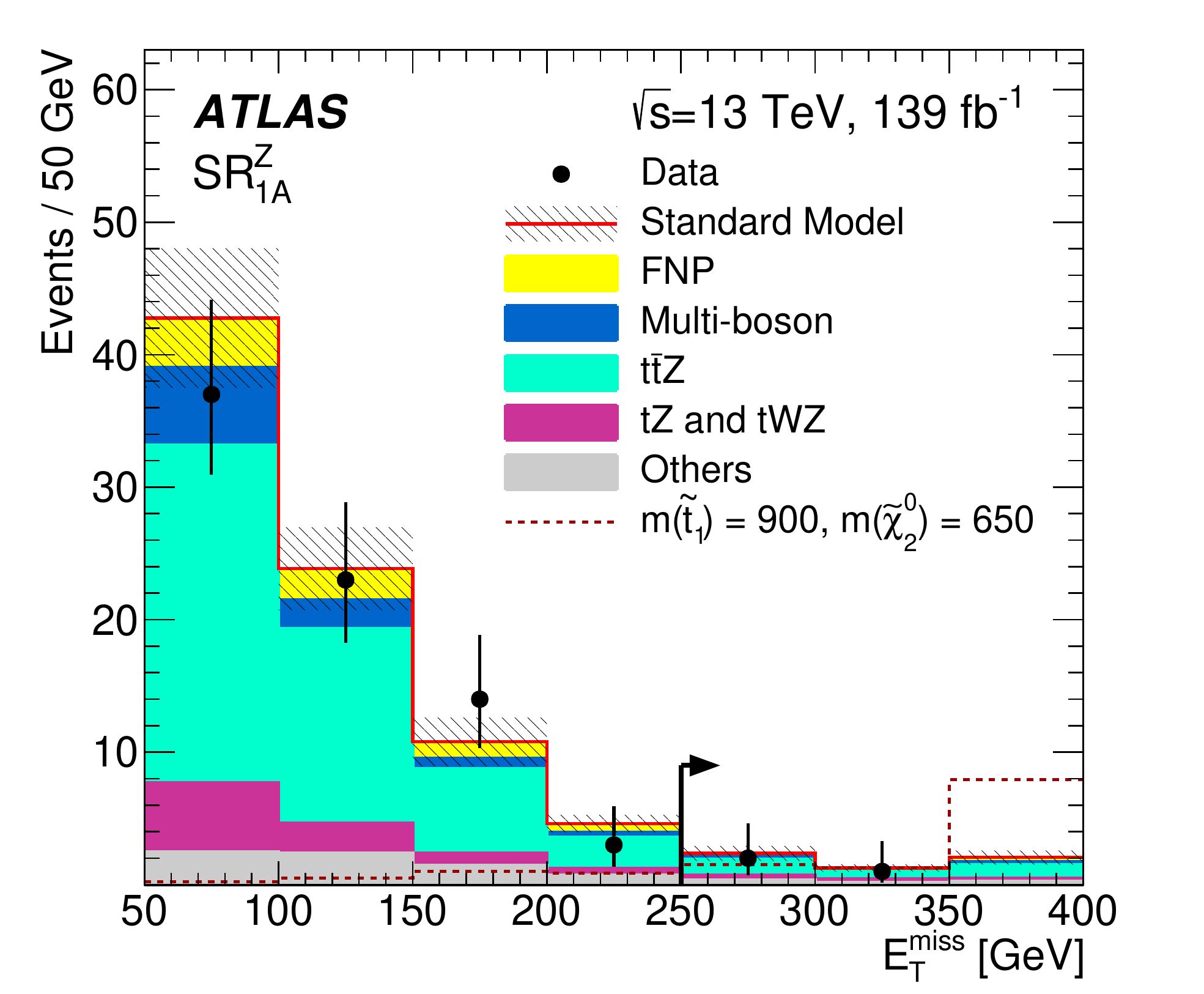}
\caption{}
\end{subfigure}
\begin{subfigure}[t]{0.4\textwidth}
\centering
\includegraphics[width=\textwidth]{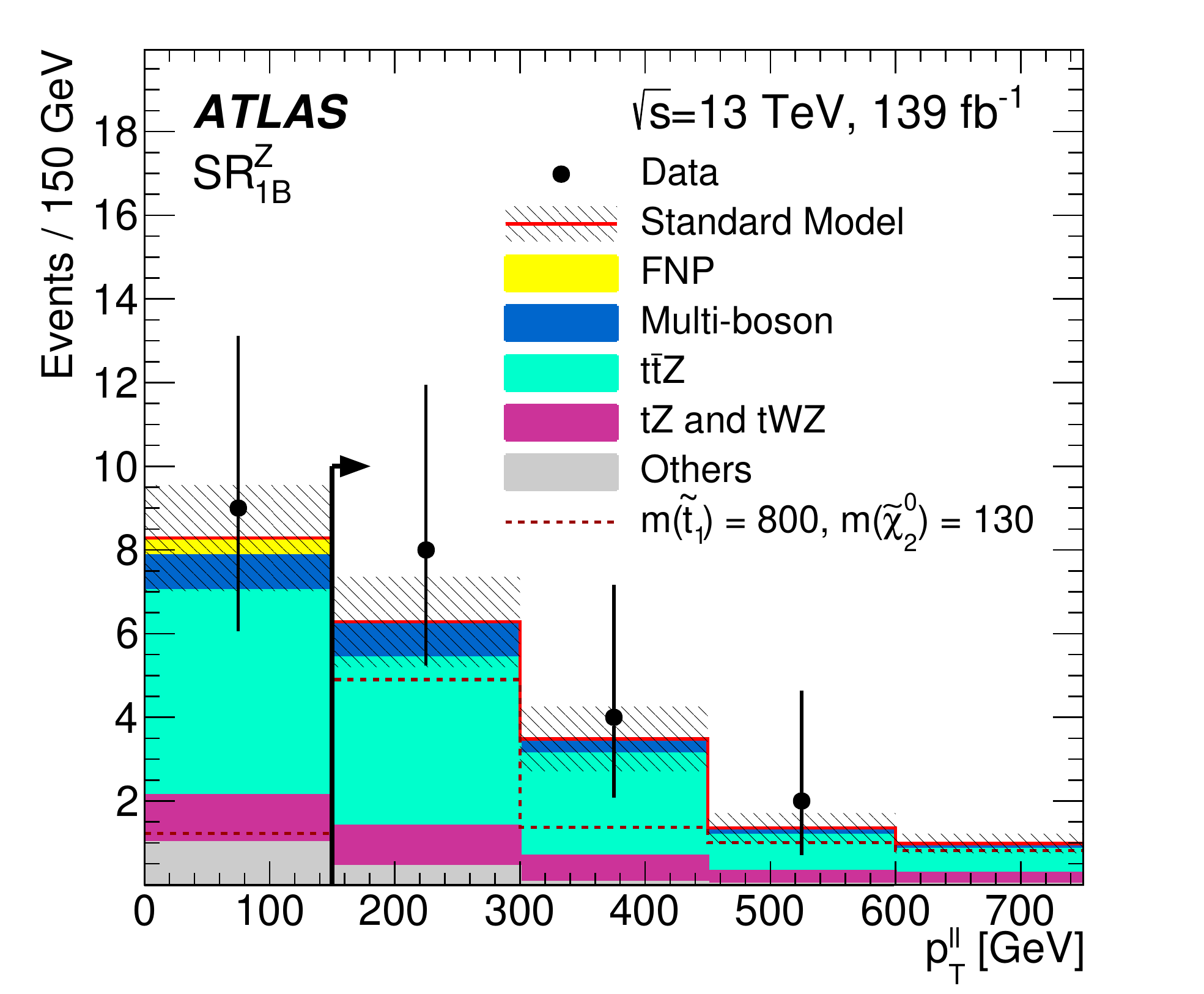}
\caption{}
\end{subfigure} \\
 
\begin{subfigure}[t]{0.4\textwidth}
\centering
\includegraphics[width=\textwidth]{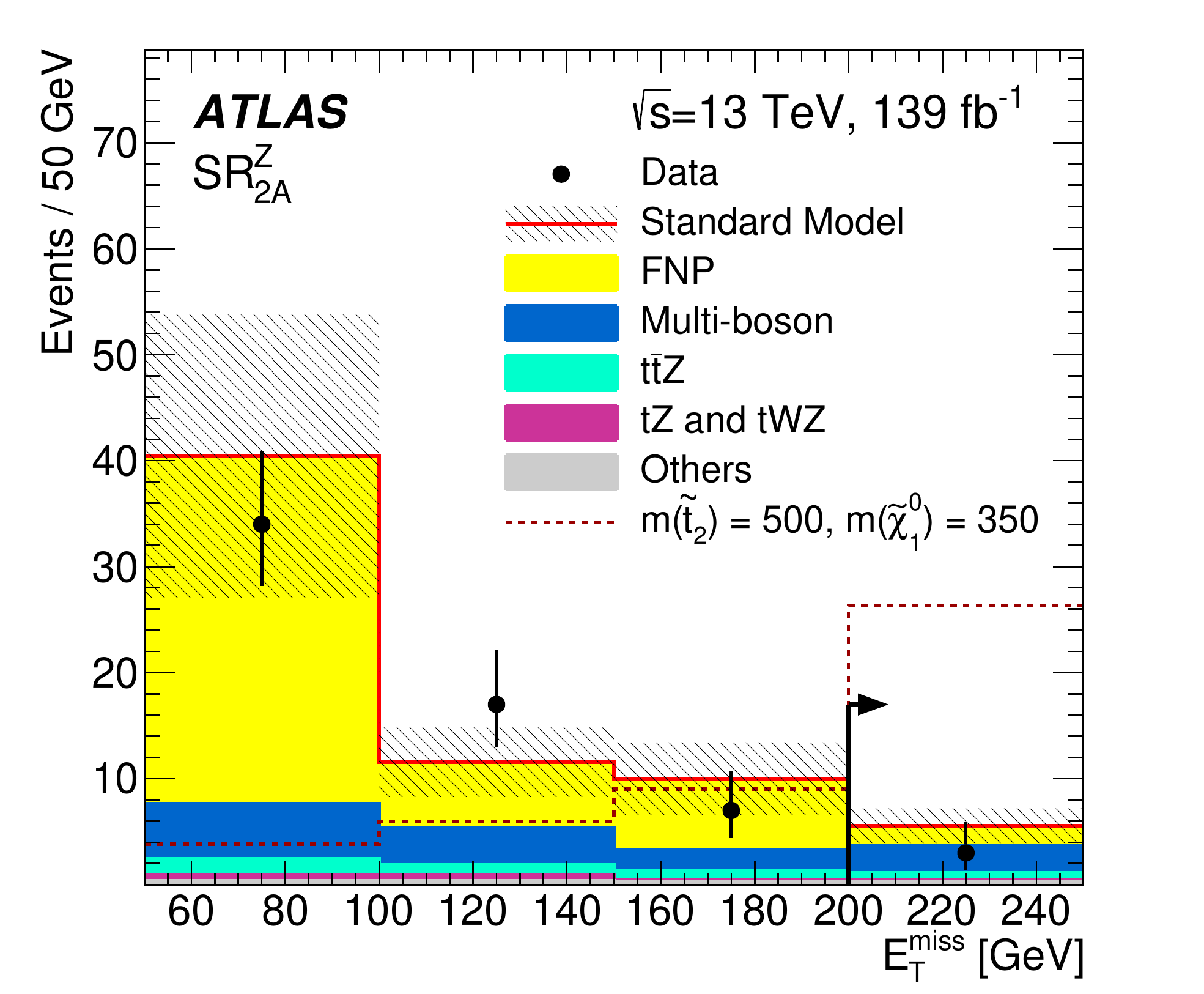}
\caption{}
\end{subfigure}
\begin{subfigure}[t]{0.4\textwidth}
\centering
\includegraphics[width=\textwidth]{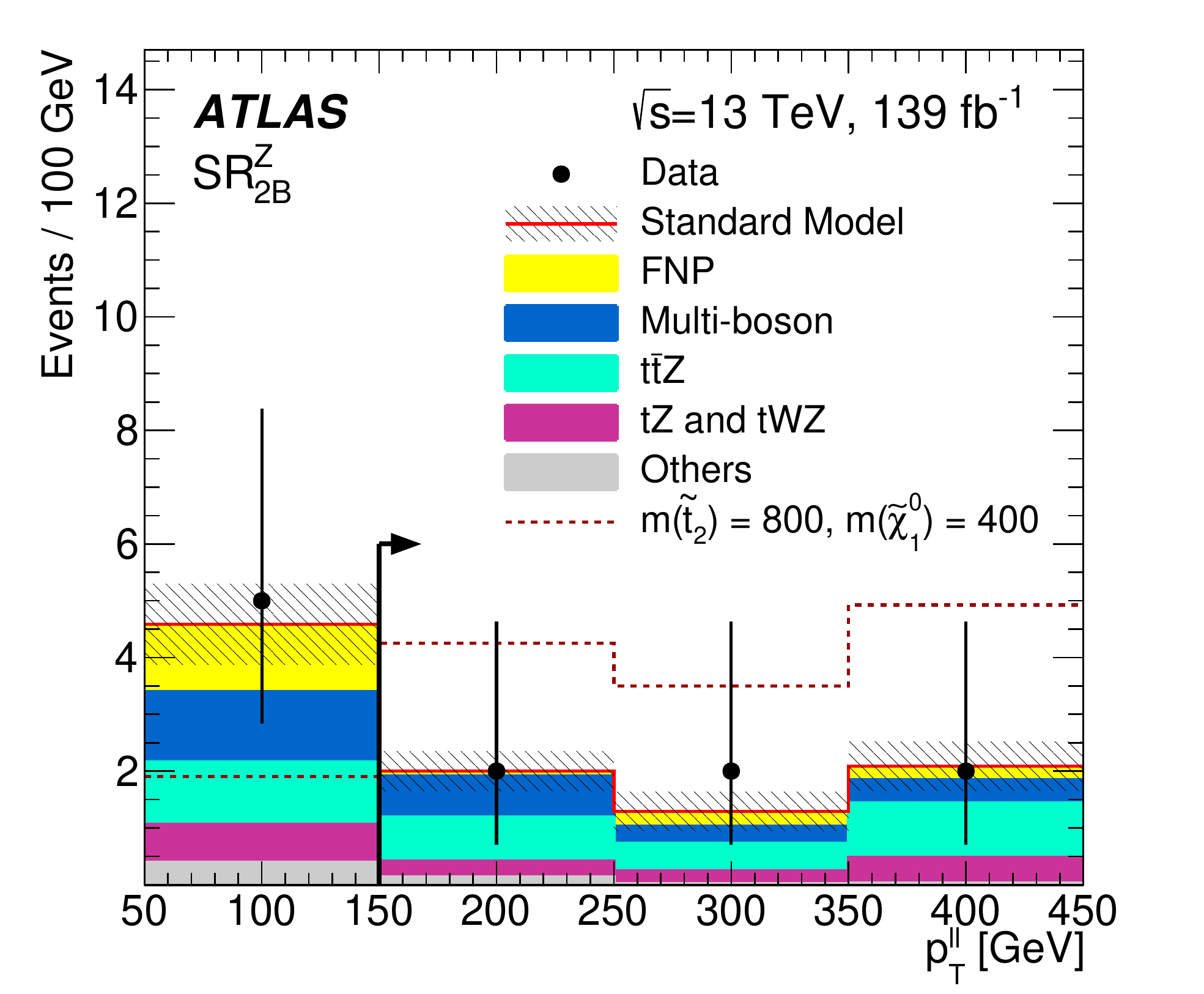}
\caption{}
\end{subfigure} \\
\end{center}
\caption{Distributions of (a) \MET\ in \SRH, (b) $p_{\textrm T}^{\ell\ell}$ in \SRL, (c) \MET\ in \SRSC, and (d) $p_{\textrm T}^{\ell\ell}$ in \SRSB\ for events passing all the SR requirements except those on the variable being plotted (the requirements are indicated by the arrows).
The contributions from all SM backgrounds are shown after the background fit described in Section~\ref{sec:bkg}; the hashed bands represent the total uncertainty.
The `Others' category contains the contributions from $\ttbar h$, $\ttbar W$, $\ttbar WW$, $\ttbar t$, $\ttbar \ttbar$, $Wh$, and $Zh$ production.
The `FNP' category represents the background from fake or non-prompt leptons.
The expected distributions for selected signal models are also shown as dashed lines.
The last bin in each figure contains the overflow.
}
\label{fig:kin_SR}
\end{figure}
 
\begin{figure}[htb!]
\begin{center}
\begin{subfigure}[t]{0.4\textwidth}
\centering
\includegraphics[width=\textwidth]{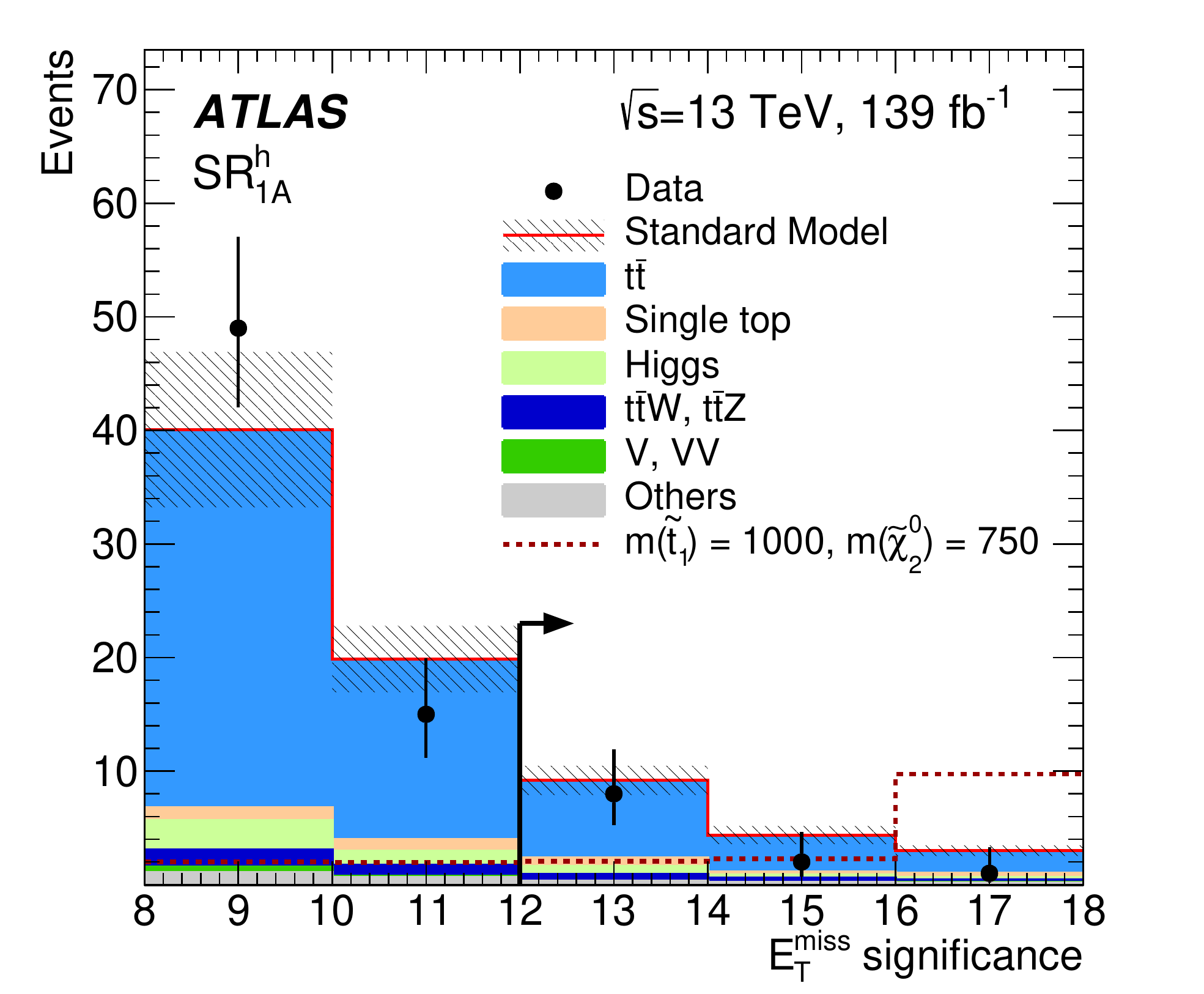}
\caption{}
\end{subfigure}
\begin{subfigure}[t]{0.4\textwidth}
\centering
\includegraphics[width=\textwidth]{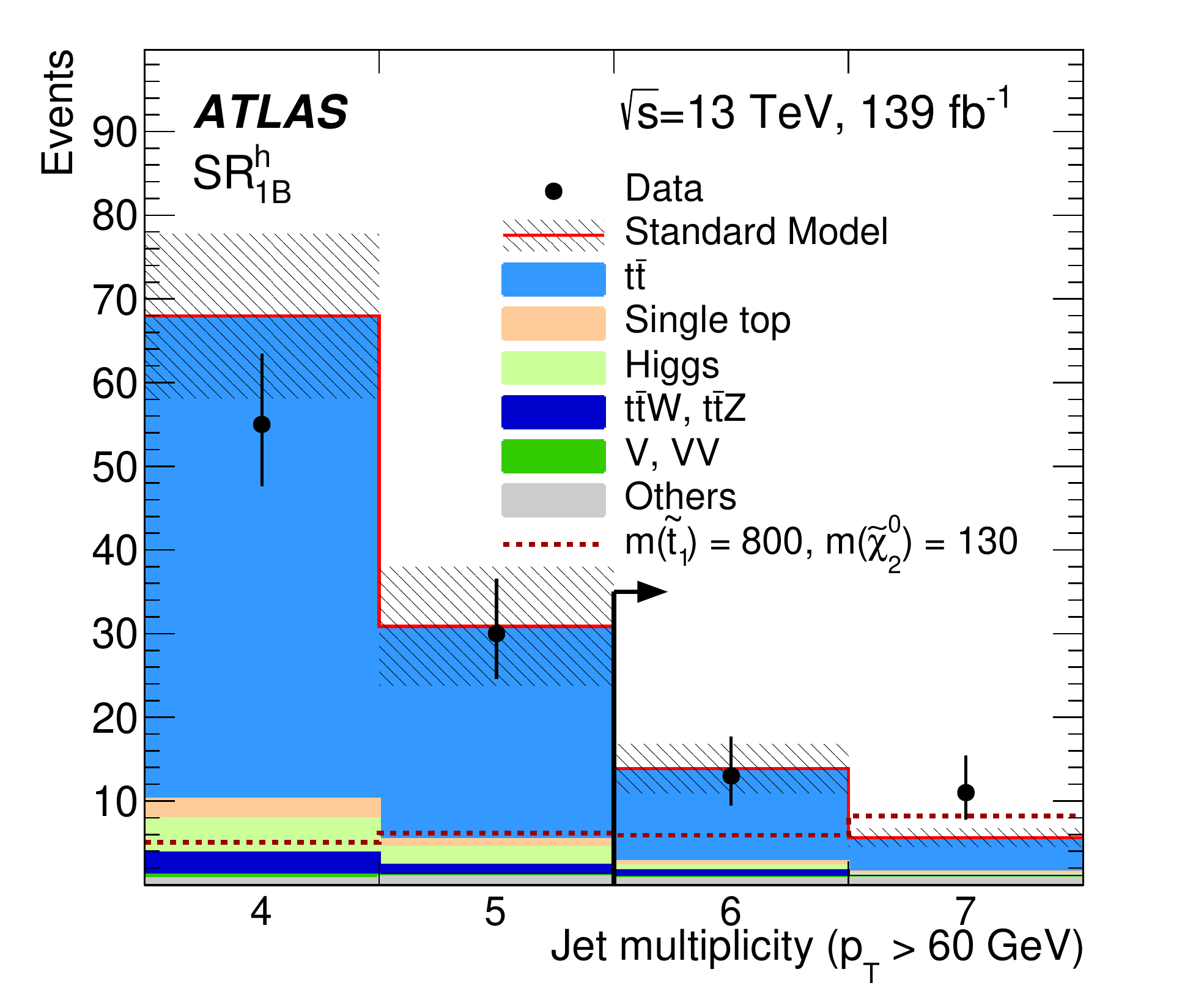}
\caption{}
\end{subfigure} \\
\end{center}
\vspace{-0.7cm}
\caption{Distributions of (a) \MET\ significance in \SRbH\ and (b) jet multiplicity in \SRbL, for events passing all the SR requirements except those on the variable being plotted (the requirements are indicated by the arrows).
The contributions from all SM backgrounds are shown after the background fit described in Section~\ref{sec:bkg}; the hashed bands represent the total uncertainty.
The `Higgs' category contains the contributions from gluon--gluon fusion, vector-boson fusion, $Wh$, $Zh$ and $\ttbar h$ production.
The `Others' category contains the contributions from $\ttbar WW$, $\ttbar WZ$, $\ttbar t$ and $\ttbar \ttbar$ production.
The expected distributions for selected signal models are also shown as dashed lines.
The last bin in each figure contains the overflow.
}
\label{fig:kin_SRH}
\end{figure}
 
\begin{table}
\caption{Observed and expected numbers of events in the 3$\ell$ signal regions. The pre-fit predictions from MC simulation are given for comparison for those backgrounds ($\ttbar Z$, multi-boson) that are normalised to data in dedicated control regions.
The `Others' category is dominated by $\ttbar W$ production and also contains the contributions from $\ttbar h$, $\ttbar WW$, $\ttbar t$, $\ttbar \ttbar$, $Wh$, and $Zh$ production.
Combined statistical and systematic uncertainties are given. The table also includes model-independent 95\% CL upper limits on the visible number of BSM events ($S^{95}_{\textrm{obs}}$), the number of BSM events given the expected number of background events ($S^{95}_{\textrm{exp}}$) and the visible BSM cross-section ($\sigma_{\mathrm{vis}}$), as well as the discovery $p$-value ($p_0$) for the background-only hypothesis, all calculated from pseudo-experiments.
The value of $p_0$ is capped at 0.5 if the observed number of events is below the expected number of events.
\label{tab:results}
}
\vspace{-0.3cm}
\begin{center}
\setlength{\tabcolsep}{0.0pc}
{\small
\sisetup{retain-explicit-plus=true}
\def\arraystretch{1.2}
\begin{tabular*}{ 0.9\textwidth}{@{\extracolsep{\fill}}l rcl c rcl c rcl c rcl}
\noalign{\smallskip}\hline\noalign{\smallskip}
& \multicolumn{3}{c}{\SRH}                        & & \multicolumn{3}{c}{\SRL}                           & & \multicolumn{3}{c}{\SRSC}                          & & \multicolumn{3}{c}{\SRSB}   \\[-0.05cm]
\noalign{\smallskip}\hline\noalign{\smallskip}
Observed events                     & \multicolumn{3}{c}{$3$}                         & & \multicolumn{3}{c}{$14$}                           & & \multicolumn{3}{c}{$3$}                            & & \multicolumn{3}{c}{$6$}     	\\
\noalign{\smallskip}\hline\noalign{\smallskip}
Total (post-fit) SM events          & $5.7$     & $\pm$ & $1.0$        				  & & $12.1$   & $\pm$ & $2.0$                           & & $5.6$  & $\pm$ & $1.6$  							& & $5.5$  & $\pm$ & $0.9$   \\
\noalign{\smallskip}\hline\noalign{\smallskip}
Post-fit, multi-boson               & $0.49$    & $\pm$ & $0.22$       				  & & $1.5$    & $\pm$ & $0.5$                           & & $2.6$  & $\pm$ & $1.0$  							& & $1.4$  & $\pm$ & $0.6$\\
Post-fit, \ttZ                      & $2.8$     & $\pm$ & $0.9$        				  & & $7.9$    & $\pm$ & $1.9$                           & & $0.70$ & $\pm$ & $0.23$ 							& & $2.2$  & $\pm$ & $0.7$\\
Fake or non-prompt leptons          & $0.74$    & $\pm$ & $0.24$       				  & & $0.04$   & $\pm$ & $0.02$                          & & $1.8$  & $\pm$ & $1.1$  							& & $0.65$ & $\pm$ & $0.12$\\
$tZ$, $tWZ$                         & $0.8$     & $\pm$ & $0.4$        				  & & $2.2$    & $\pm$ & $1.2$                           & & $0.19$ & $\pm$ & $0.10$ 							& & $1.0$  & $\pm$ & $0.5$\\
Others                              & $0.84$    & $\pm$ & $0.18$       				  & & $0.51$   & $\pm$ & $0.11$                          & & $0.25$ & $\pm$ & $0.07$ 							& & $0.19$ & $\pm$ & $0.04$\\
\noalign{\smallskip}\hline\noalign{\smallskip}
Pre-fit, multi-boson                & $0.61$    & $\pm$ & $0.23$       				  & & $1.9$    & $\pm$ & $0.5$                           & & $3.3$  & $\pm$ & $0.9$  							& & $1.8$  & $\pm$ & $0.7$\\
Pre-fit, \ttZ                       & $3.0$     & $\pm$ & $0.7$        				  & & $8.5$    & $\pm$ & $1.6$                           & & $0.76$ & $\pm$ & $0.21$ 							& & $2.4$  & $\pm$ & $0.5$ \\
\noalign{\smallskip}\hline\noalign{\smallskip}
$S_{\textrm{obs}}^{95}$           & \multicolumn{3}{c}{4.5} 					      & & \multicolumn{3}{c}{11.7}					         &  & \multicolumn{3}{c}{4.9} 				            & & \multicolumn{3}{c}{7.0} \\
$S_{\textrm{exp}}^{95}$           & \multicolumn{3}{c}{${ 6.2 }^{ +2.6 }_{ -1.6 }$}   & & \multicolumn{3}{c}{$ { 9.2 }^{ +4.1 }_{ -1.3 }$}	 &  & \multicolumn{3}{c}{$ { 6.1 }^{ +2.6 }_{ -1.7 }$}  & & \multicolumn{3}{c}{$ { 6.6 }^{ +2.5 }_{ -1.8 }$} \\
$\sigma_{\mathrm{vis}}$ [fb]      & \multicolumn{3}{c}{0.03}					      & & \multicolumn{3}{c}{0.08} 				             & 	& \multicolumn{3}{c}{0.03} 					        & & \multicolumn{3}{c}{0.05} \\
$p_0$                             & \multicolumn{3}{c}{0.50}						  & & \multicolumn{3}{c}{0.31} 					         &  & \multicolumn{3}{c}{0.50} 					        & & \multicolumn{3}{c}{0.4} \\
\noalign{\smallskip}\hline\noalign{\smallskip}
\end{tabular*}
}
\end{center}
\end{table}
 
\FloatBarrier

\begin{table}
\caption{Observed and expected numbers of events in the 1$\ell$ signal regions. The pre-fit predictions from MC simulation are given for comparison for the $\ttbar$ background that is normalised to data in a dedicated control region.
The `Higgs' category contains the contributions from gluon--gluon fusion, vector-boson fusion, $Wh$, $Zh$ and $\ttbar h$ production.
The `Others' category contains the contributions from $\ttbar WW$, $\ttbar WZ$, $\ttbar t$ and $\ttbar \ttbar$ production.
Combined statistical and systematic uncertainties are given. The table also includes model-independent 95\% CL upper limits on the visible number of BSM events ($S^{95}_{\textrm{obs}}$), the number of BSM events given the expected number of background events ($S^{95}_{\textrm{exp}}$) and the visible BSM cross-section ($\sigma_{\mathrm{vis}}$), as well as the discovery $p$-value ($p_0$) for the background-only hypothesis, all calculated from pseudo-experiments.
The value of $p_0$ is capped at 0.5 if the observed number of events is below the expected number of events.
\label{tab:resultsH}
}
\begin{center}
\setlength{\tabcolsep}{0.0pc}
{\small
\def\arraystretch{1.2}
\begin{tabular*}{0.51 \textwidth}{@{\extracolsep{\fill}}l rcl c rcl}
\noalign{\smallskip}\hline\noalign{\smallskip}
& \multicolumn{3}{c}{\SRbH}            			   &       & \multicolumn{3}{c}{\SRbL}           \\[-0.05cm]
\noalign{\smallskip}\hline\noalign{\smallskip}
Observed events                             & \multicolumn{3}{c}{$11$}             	           &       & \multicolumn{3}{c}{$24$}     	\\
\noalign{\smallskip}\hline\noalign{\smallskip}
Total (post-fit) SM events                  & $17$                                             & $\pm$ & $3$       & & $19$   & $\pm$ & $5$   \\
\noalign{\smallskip}\hline\noalign{\smallskip}
Post-fit, $\ttbar$                          & $12$                                             & $\pm$ & $3$       & & $15$   & $\pm$ & $5$\\
$V$, $VV$                                   & $0.05$                                           & $\pm$ & $0.05$    & & $0.13$ & $\pm$ & $ 0.08$              \\
$\ttbar W$, $\ttbar Z$                      & $1.16$                                           & $\pm$ & $0.26$    & & $0.95$ & $\pm$ & $0.25$              \\
Higgs                                       & $1.19$                                           & $\pm$ & $0.21$    & & $0.9$  & $\pm$ & $0.4$              \\
Single top                                  & $1.38$                                           & $\pm$ & $0.23$    & & $0.74$ & $\pm$ & $0.22$              \\
Others                                      & $0.68$                                           & $\pm$ & $0.13$    & & $1.53$ & $\pm$ & $0.32$              \\
\noalign{\smallskip}\hline\noalign{\smallskip}
Pre-fit, $\ttbar$                           & $11.0$                                           & $\pm$ & $2.4$     & & $14$   & $\pm$ & $4$ \\
\noalign{\smallskip}\hline\noalign{\smallskip}
$S_{\textrm{obs}}^{95}$                     & \multicolumn{3}{c}{7.0} 					       &       & \multicolumn{3}{c}{18.1}			 \\
$S_{\textrm{exp}}^{95}$                     & \multicolumn{3}{c}{${10.3}^{+4.4 }_{ -3.2 }$}	   &       & \multicolumn{3}{c}{$ {14.2}^{ +6.0 }_{ -3.8 }$}	 \\
$\sigma_{\mathrm{vis}}$ [fb]                & \multicolumn{3}{c}{0.05} 						   &       & \multicolumn{3}{c}{0.13}					 \\
$p_0$                                     	& \multicolumn{3}{c}{0.50}						   &       & \multicolumn{3}{c}{0.25}					 \\
\noalign{\smallskip}\hline\noalign{\smallskip}
\end{tabular*}
}
\end{center}
\end{table}
 
\FloatBarrier

\begin{table}[hbt!]
\caption{\label{tab:SRexcl} Selection criteria used in the shape-fit to derive the model-dependent exclusion limits. The additional SR labelled as SR$_{1{\textrm{AB}}}^{h}$ overlaps with both \SRbH\ and \SRbL\ defined in Table~\ref{tab:SRdefH}.}
\def\arraystretch{1.2}
\begin{center}
\begin{tabular}{l l l}
\noalign{\smallskip}\hline\noalign{\smallskip}
 
\SRH & $\met$ [\GeV] & 200--250, 250--300, 300--350, $>$350 \\\hline
\SRL & $p_{\textrm T}^{\ell\ell}$ [\GeV] & 150--300, 300--450, 450--600, $>$600 \\\hline
\SRSB & $\met$ [\GeV] & 300--350, $>$350 \\
& $p_{\textrm T}^{\ell\ell}$ [\GeV] & 50--150, $>$150 \\\hline
\SRbH  & $n_{h\textrm{-cand}}$                            & 1, $\geq$2  \\
& $n_{\textrm{jets}}$ ($\pt>60$~\GeV)   & 4 \\
& $\cal S$                                  & 10--12, 12--14   \\\hline
\SRbL  & $n_{h\textrm{-cand}}$                            & 1, $\geq$2  \\
& $n_{\textrm{jets}}$ ($\pt>60$~\GeV)   & 5, $\geq$6 \\
& $\cal S$                                    & 7--14   \\\hline
SR$_{1{\textrm{AB}}}^{h}$  & $n_{h\textrm{-cand}}$             & $\geq$1  \\
& $n_{\textrm{jets}}$ ($\pt>60$~\GeV)   & $\geq$4 \\
& $\cal S$                                    & $\geq$14   \\\hline
\end{tabular}
\end{center}
\end{table}
 
\FloatBarrier
 
\begin{figure}[htb!]
\begin{center}
\includegraphics[width=\textwidth]{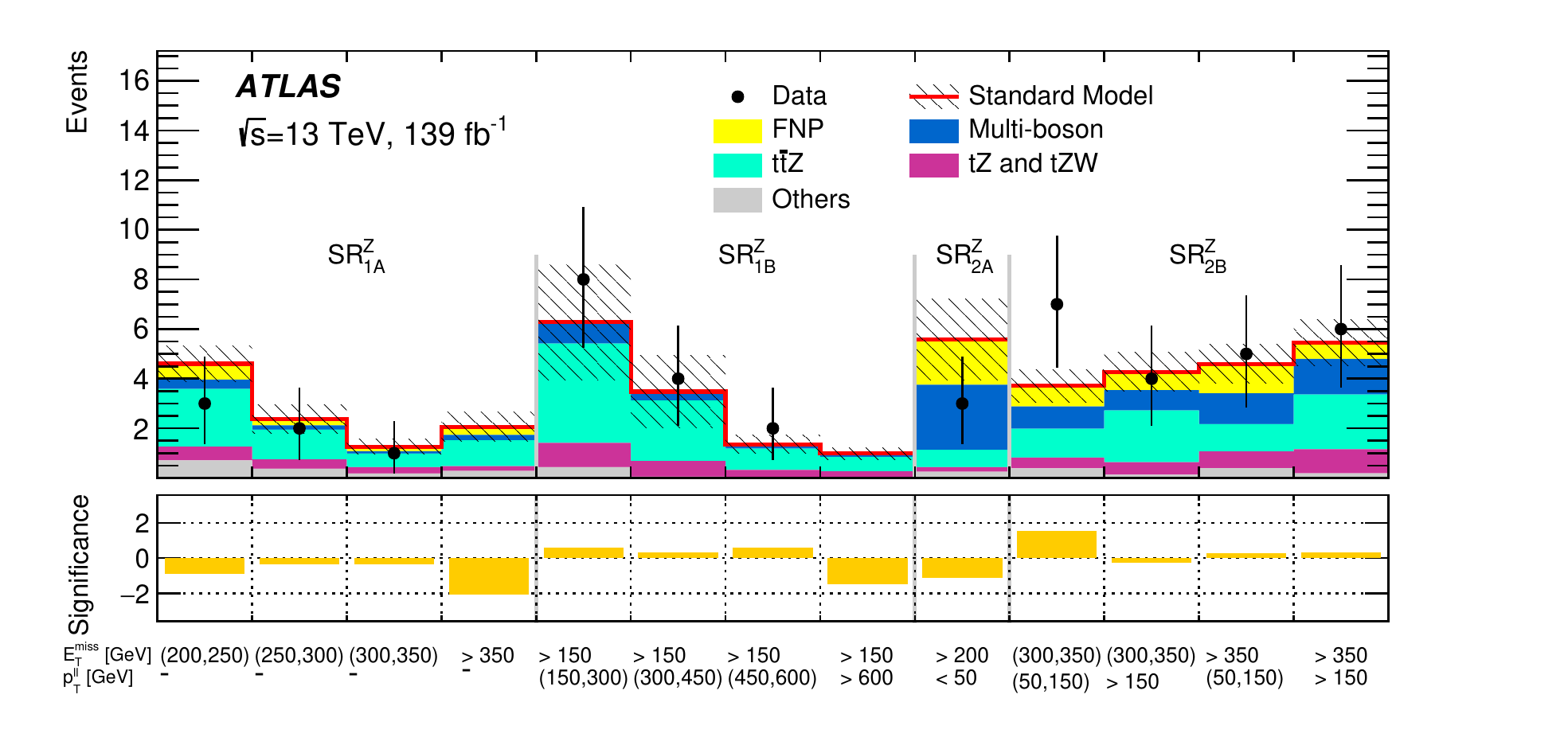}
\caption{Comparison of the observed and expected event yields in all the 3$\ell$ SRs and bins used for the model-dependent exclusion limits.
The `Others' category contains the contributions from $\ttbar h$, $\ttbar W$, $\ttbar WW$, $\ttbar t$, $\ttbar \ttbar$, $Wh$, and $Zh$ production.
The `FNP' category represents the background from fake or non-prompt leptons.
The lower panel shows the significance in each SR bin, computed as described in Ref.~\cite{RooFit1}.}
\label{fig:SRsumm}
\end{center}
\end{figure}

\begin{figure}[htb!]
\begin{center}
\includegraphics[width=\textwidth]{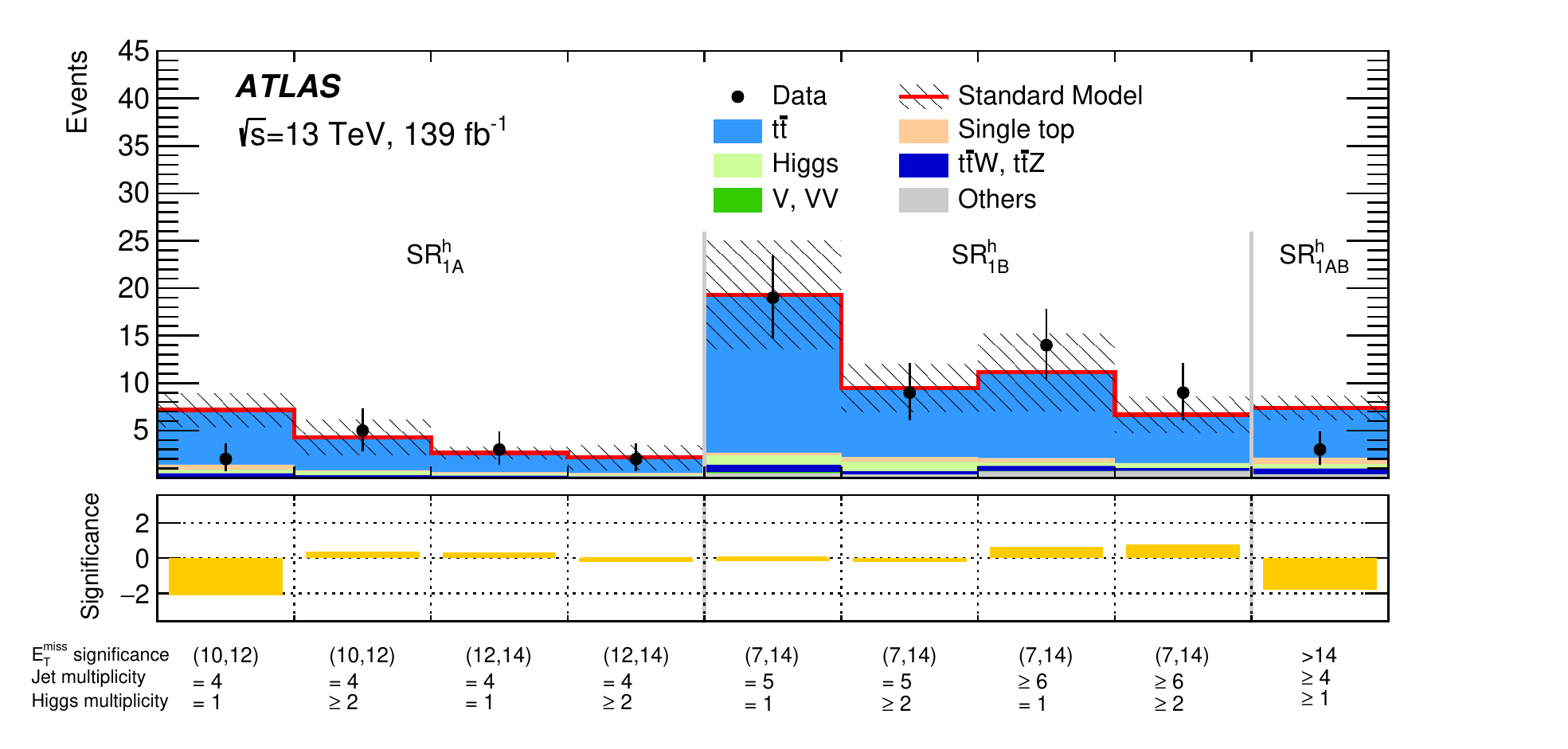}
\caption{Comparison of the observed and expected event yields in all the 1$\ell$ SRs and bins used for the model-dependent exclusion limits.
The `Higgs' category contains the contributions from gluon--gluon fusion, vector-boson fusion, $Wh$, $Zh$ and $\ttbar h$ production.
The `Others' category contains the contributions from $\ttbar WW$, $\ttbar WZ$, $\ttbar t$ and $\ttbar \ttbar$ production.
The lower panel shows the significance in each SR bin, computed as described in Ref.~\cite{RooFit1}.}
\label{fig:SRHsumm}
\end{center}
\end{figure}
 
\FloatBarrier
 
Figure~\ref{fig:LimitStop1} shows the exclusion limits in the $\stopone\to t\ninotwo$ with $\ninotwo\to h/Z\ninoone$ simplified model with 50\% branching ratios to each $\ninotwo$ decay mode.
These results are obtained from the statistical combination of the shape-fit bins of \SRH,  \SRL, \SRbH, \SRbL\ and SR$_{1{\textrm{AB}}}^{h}$ shown in Table~\ref{tab:SRexcl}.
The bins of \SRbH, \SRbL\ and SR$_{1{\textrm{AB}}}^{h}$ are separately combined with the bins of \SRH\ and  \SRL.
For each combination of sparticle masses, only the option among those two with best expected sensitivity is considered for the final limit setting.
The change in best expected combination is responsible for the kink at \stopone masses of about 900--1000~\GeV\  and \ninotwo masses below 200~\GeV.
For \ninotwo masses above 225~\GeV, \stopone masses up to about 1100~\GeV\ are excluded at 95\% CL, while \stopone masses below 1220~\GeV\ are excluded for a \ninotwo mass of 925~\GeV.
These results improve upon the existing limits on the \stopone mass in this model by approximately 300~\GeV~\cite{SUSY-2016-20}.
 
The same statistical combination strategy is also used to obtain exclusion limits in the $\stopone\to t\ninotwo$ with $\ninotwo\to h/Z\ninoone$ simplified model for different $\ninotwo\to h/Z\ninoone$ decay branching ratios, shown in Figure~\ref{fig:LimitStop1BR}.
For \ninotwo masses above 300~\GeV, the exclusion limits on  \stopone masses vary by at most 40~\GeV\ depending on the $\ninotwo\to h/Z\ninoone$ branching ratios.
However, for \ninotwo masses below 200~\GeV\ the exclusion limits on \stopone masses are up to 300~\GeV\ better for models featuring only $\ninotwo\to Z\ninoone$ decays compared with models only considering the $\ninotwo\to h\ninoone$ decays.
 
\begin{figure}[htb!]
\begin{center}
\includegraphics[width=0.68\textwidth]{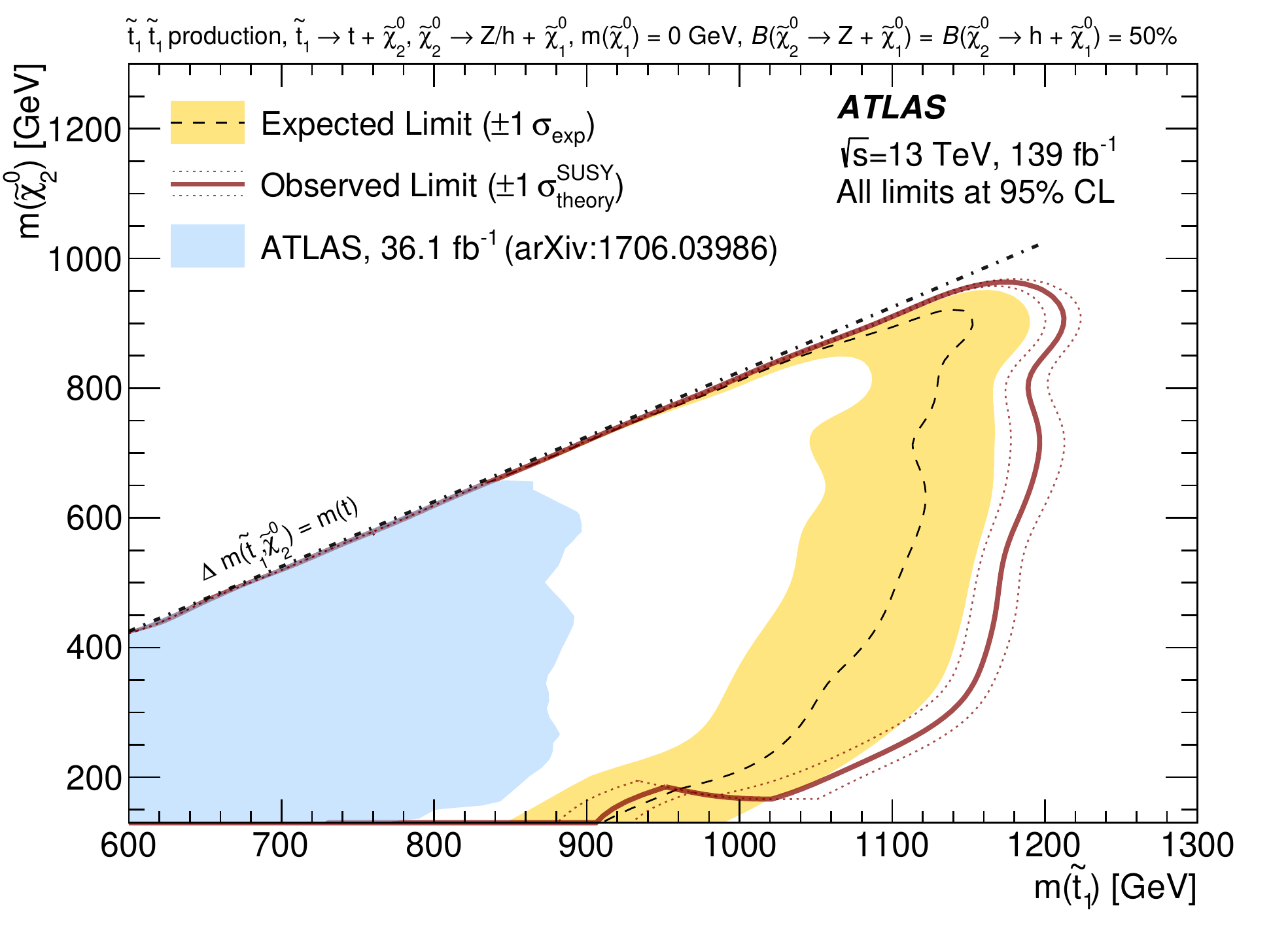}
\caption{Exclusion limits at 95\% CL on the masses of the $\stopone$ and $\ninotwo$, for a fixed $m(\ninoone) = 0$~\GeV, assuming ${\cal B}(\ninotwo \rightarrow Z\ninoone) = 0.5$ and ${\cal B}(\ninotwo \rightarrow h\ninoone) = 0.5$.
The dashed line and the shaded band are the expected limit and its $\pm1 \sigma$ uncertainty, respectively. The thick solid line is the observed limit for the central value of the signal cross-section.
The expected and observed limits do not include the effect of the theoretical uncertainties in the signal cross-section. The dotted lines show the effect on the observed limit of varying the signal cross-section by $\pm 1 \sigma$ of the theoretical uncertainty.
Results are compared with the observed limits obtained by the previous ATLAS search in Ref.~\cite{SUSY-2016-20}.}
\label{fig:LimitStop1}
\end{center}
\end{figure}
 
\begin{figure}[htb!]
\begin{center}
\includegraphics[width=0.68\textwidth]{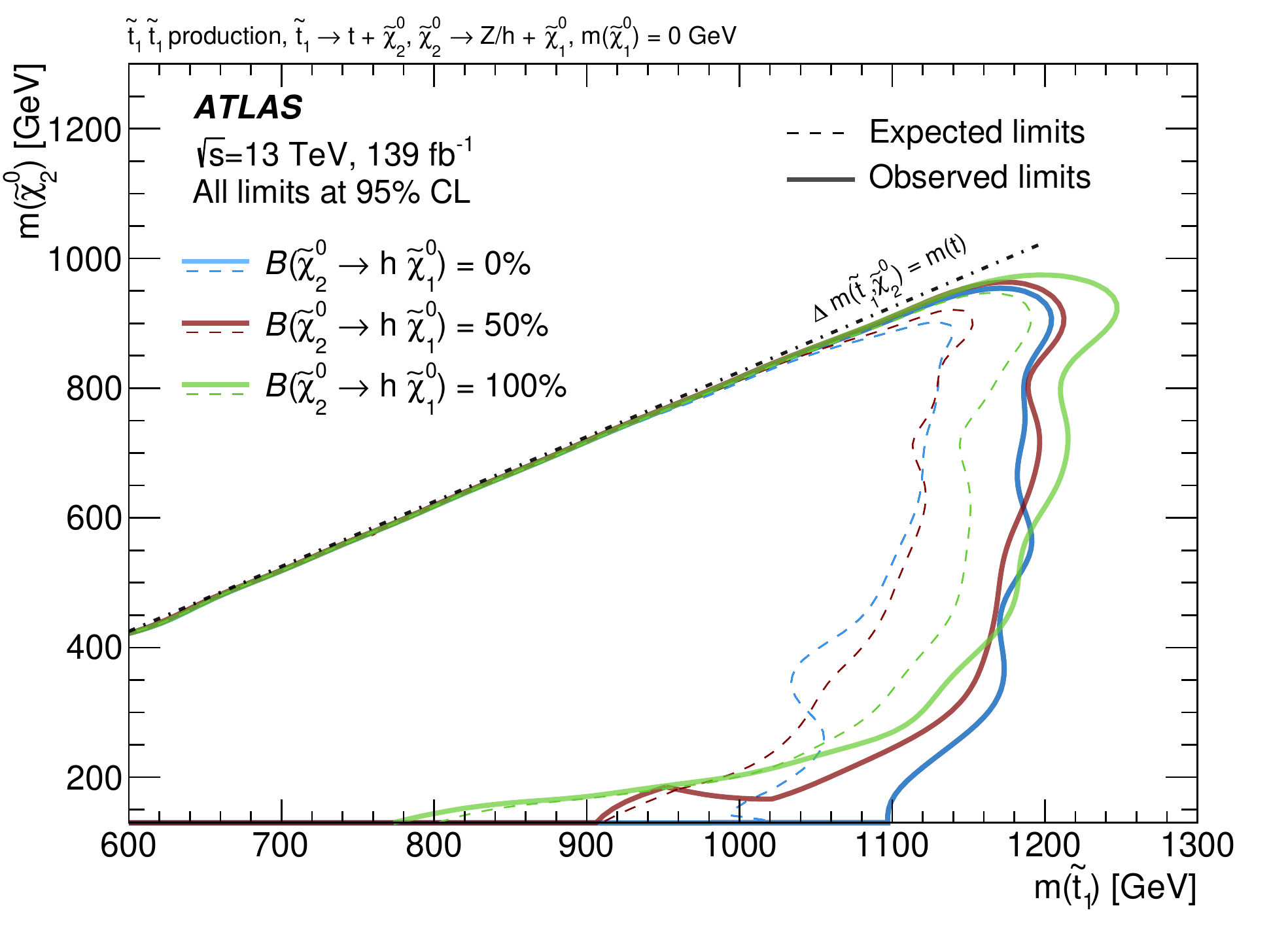}
\caption{Exclusion limits at 95\% CL on the masses of the $\stopone$ and $\ninotwo$, for a fixed $m(\ninoone) = 0$~\GeV\ and different values of ${\cal B}(\ninotwo \rightarrow h\ninoone)$ with ${\cal B}(\ninotwo \rightarrow Z\ninoone) = 1 - {\cal B}(\ninotwo \rightarrow h\ninoone)$.
The dashed and solid lines are the expected and observed limits for the central value of the signal cross-section, respectively.
The expected and observed limits do not include the effect of the theoretical uncertainties in the signal cross-section.
}
\label{fig:LimitStop1BR}
\end{center}
\end{figure}
 
\FloatBarrier
 
Figure~\ref{fig:LimitStop2} shows the exclusion limits in the $\stoptwo\to Z \stopone$ with $\stopone \to bff'\ninoone$ simplified model for a mass difference between the $\stopone$ and $\ninoone$ of 40~\GeV. Several options for the $\stopone$ and $\ninoone$ mass difference are also considered and the sensitivity is found to appreciably decrease only for values below 20~\GeV.
These results are obtained from the statistical combination of \SRSC\ and the shape-fit bins of \SRSB\ shown in Table~\ref{tab:SRexcl}.
The shape of the contour is driven by \SRSC\ and \SRSB\ being most sensitive to small and large mass splittings between the \stoptwo and the \ninoone respectively.
Masses of the \stoptwo up to 875~\GeV\ are excluded at 95\% CL for a \ninoone mass of about 350~\GeV\, and \ninoone masses of approximately 520 (450)~\GeV\ are excluded for \stoptwo masses of 650 (800)~\GeV, extending beyond the previous limits on the \ninoone mass from Ref.~\cite{SUSY-2016-17} by up to 160~\GeV.
 
\begin{figure}[htb!]
\begin{center}
\includegraphics[width=0.68\textwidth]{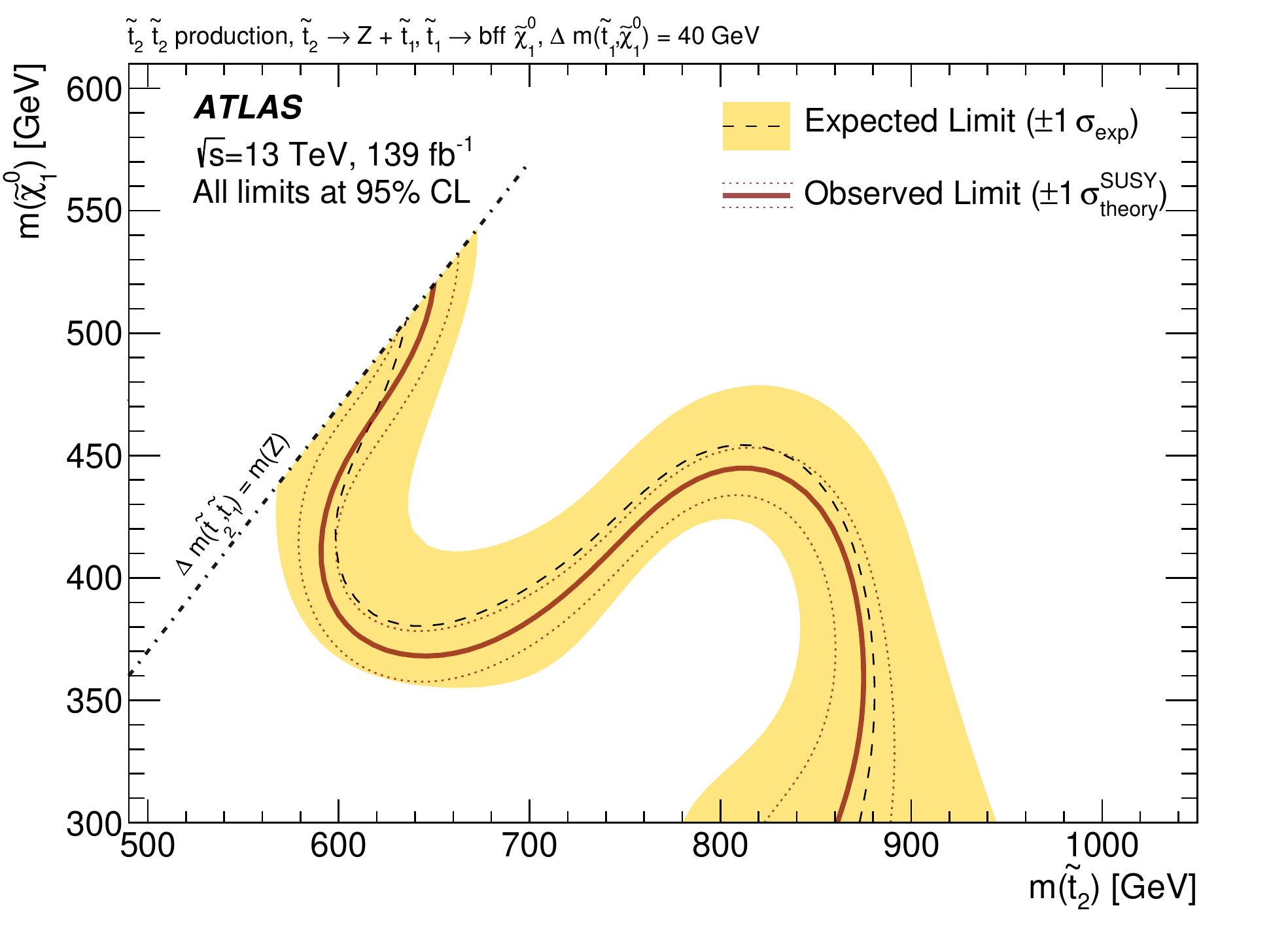}
\caption{Exclusion limits at 95\% CL on the masses of the $\stoptwo$ and $\ninoone$,
for a fixed $m(\stopone)-m(\ninoone) = 40$~\GeV\ and assuming ${\cal B}(\stoptwo \rightarrow Z\stopone) = 1$.
The dashed line and the shaded band are the expected limit and its $\pm1 \sigma$ uncertainty, respectively. The thick solid line is the observed limit for the central value of the signal cross-section.
The expected and observed limits do not include the effect of the theoretical uncertainties in the signal cross-section. The dotted lines show the effect on the observed limit of varying the signal cross-section by $\pm 1 \sigma$ of the theoretical uncertainty.
}
\label{fig:LimitStop2}
\end{center}
\end{figure}
 
\FloatBarrier
 
\section{Conclusion}
\label{sec:conclusion}
 
A search for direct top squark pair production in events with a leptonically decaying $Z$ boson or a SM Higgs boson decaying into a $b$-quark pair is presented. The analysis uses 139~\ifb\ of proton--proton collision data at $\sqrt{s} = 13$~\TeV\ recorded by the ATLAS detector at the LHC.
No excess over the SM background predictions is observed, and exclusion limits are presented for a selection of simplified models.
The limits exclude, at 95\% confidence level, $\stopone$ masses up to 1220~\GeV\ in models featuring $\stopone$ production and $\stopone \to t\ninotwo$ with $\ninotwo \to Z/h\ninoone$ decays, and $\stoptwo$ masses up to 875~\GeV\ in models featuring $\stoptwo$ production and $\stoptwo\to Z\stopone$ with $\stopone \to b ff'\ninoone$ decays.
Compared with previous limits, these results extend the mass parameter space exclusion by up to 300~\GeV\ in $\stopone$ mass and by up to 160~\GeV\ in $\ninoone$ mass in the considered $\stoptwo$ model.

\section*{Acknowledgements}
 
% The next lines are included from the .//acknowledgements/Acknowledgements.tex input file

We thank CERN for the very successful operation of the LHC, as well as the
support staff from our institutions without whom ATLAS could not be
operated efficiently.
 
We acknowledge the support of ANPCyT, Argentina; YerPhI, Armenia; ARC, Australia; BMWFW and FWF, Austria; ANAS, Azerbaijan; SSTC, Belarus; CNPq and FAPESP, Brazil; NSERC, NRC and CFI, Canada; CERN; CONICYT, Chile; CAS, MOST and NSFC, China; COLCIENCIAS, Colombia; MSMT CR, MPO CR and VSC CR, Czech Republic; DNRF and DNSRC, Denmark; IN2P3-CNRS and CEA-DRF/IRFU, France; SRNSFG, Georgia; BMBF, HGF and MPG, Germany; GSRT, Greece; RGC and Hong Kong SAR, China; ISF and Benoziyo Center, Israel; INFN, Italy; MEXT and JSPS, Japan; CNRST, Morocco; NWO, Netherlands; RCN, Norway; MNiSW and NCN, Poland; FCT, Portugal; MNE/IFA, Romania; MES of Russia and NRC KI, Russia Federation; JINR; MESTD, Serbia; MSSR, Slovakia; ARRS and MIZ\v{S}, Slovenia; DST/NRF, South Africa; MINECO, Spain; SRC and Wallenberg Foundation, Sweden; SERI, SNSF and Cantons of Bern and Geneva, Switzerland; MOST, Taiwan; TAEK, Turkey; STFC, United Kingdom; DOE and NSF, United States of America. In addition, individual groups and members have received support from BCKDF, CANARIE, Compute Canada and CRC, Canada; ERC, ERDF, Horizon 2020, Marie Sk{\l}odowska-Curie Actions and COST, European Union; Investissements d'Avenir Labex, Investissements d'Avenir Idex and ANR, France; DFG and AvH Foundation, Germany; Herakleitos, Thales and Aristeia programmes co-financed by EU-ESF and the Greek NSRF, Greece; BSF-NSF and GIF, Israel; CERCA Programme Generalitat de Catalunya and PROMETEO Programme Generalitat Valenciana, Spain; G\"{o}ran Gustafssons Stiftelse, Sweden; The Royal Society and Leverhulme Trust, United Kingdom.
 
The crucial computing support from all WLCG partners is acknowledged gratefully, in particular from CERN, the ATLAS Tier-1 facilities at TRIUMF (Canada), NDGF (Denmark, Norway, Sweden), CC-IN2P3 (France), KIT/GridKA (Germany), INFN-CNAF (Italy), NL-T1 (Netherlands), PIC (Spain), ASGC (Taiwan), RAL (UK) and BNL (USA), the Tier-2 facilities worldwide and large non-WLCG resource providers. Major contributors of computing resources are listed in Ref.~\cite{ATL-SOFT-PUB-2020-001}.
 
% End of text imported from the .//acknowledgements/Acknowledgements.tex input file
 
\clearpage
\printbibliography

\clearpage % ATLAS Collaboration author list
% Reference date of SUSY-2018-21 is 2020-03-11
% Author list last updated on date 11-AUG-20
% Data extracted on 11-Aug-2020 for paper reference SUSY-2018-21
% at 5:43pm
 
\begin{flushleft}
\hypersetup{urlcolor=black}
{\Large The ATLAS Collaboration}

\bigskip

\AtlasOrcid[0000-0002-6665-4934]{G.~Aad}$^\textrm{\scriptsize 102}$,    
\AtlasOrcid[0000-0002-5888-2734]{B.~Abbott}$^\textrm{\scriptsize 128}$,    
\AtlasOrcid{D.C.~Abbott}$^\textrm{\scriptsize 103}$,    
\AtlasOrcid[0000-0002-2788-3822]{A.~Abed~Abud}$^\textrm{\scriptsize 36}$,    
\AtlasOrcid[0000-0002-1002-1652]{K.~Abeling}$^\textrm{\scriptsize 53}$,    
\AtlasOrcid[0000-0002-2987-4006]{D.K.~Abhayasinghe}$^\textrm{\scriptsize 94}$,    
\AtlasOrcid[0000-0002-8496-9294]{S.H.~Abidi}$^\textrm{\scriptsize 166}$,    
\AtlasOrcid[0000-0002-8279-9324]{O.S.~AbouZeid}$^\textrm{\scriptsize 40}$,    
\AtlasOrcid{N.L.~Abraham}$^\textrm{\scriptsize 155}$,    
\AtlasOrcid[0000-0001-5329-6640]{H.~Abramowicz}$^\textrm{\scriptsize 160}$
% ...
\bigskip
\\

% ...
$^{36}$CERN, Geneva; Switzerland.\\
% ...
$^{40}$Niels Bohr Institute, University of Copenhagen, Copenhagen; Denmark.\\
% ...
$^{53}$II. Physikalisches Institut, Georg-August-Universit\"{a}t G\"ottingen, G\"ottingen; Germany.\\
% ...
$^{94}$Department of Physics, Royal Holloway University of London, Egham; United Kingdom.\\
% ...
$^{102}$CPPM, Aix-Marseille Universit\'e, CNRS/IN2P3, Marseille; France.\\
$^{103}$Department of Physics, University of Massachusetts, Amherst MA; United States of America.\\
% ...
$^{128}$Homer L. Dodge Department of Physics and Astronomy, University of Oklahoma, Norman OK; United States of America.\\
% ...
$^{155}$Department of Physics and Astronomy, University of Sussex, Brighton; United Kingdom.\\
% ...
$^{160}$Raymond and Beverly Sackler School of Physics and Astronomy, Tel Aviv University, Tel Aviv; Israel.\\
% ...
$^{166}$Department of Physics, University of Toronto, Toronto ON; Canada.\\
% ...

\end{flushleft}

% Created with Glance <Atlas.Glance@cern.ch>

\end{document}